\documentclass[a4paper,12pt]{article}

\RequirePackage[OT1]{fontenc}
\RequirePackage{amsthm,amsmath,amssymb,amsfonts,latexsym}
\RequirePackage{natbib}
\RequirePackage{hypernat}
\usepackage{setspace}
\usepackage{graphics,graphicx,epsfig}
\usepackage[pdftex]{color}
\usepackage{bm}
\usepackage{arydshln}
\usepackage{url}
\usepackage{rotating}
\usepackage{multicol}
\usepackage{multirow}
\usepackage{rotating}
\usepackage[hang,flushmargin]{footmisc} 
\usepackage[colorlinks=true,citecolor=blue,urlcolor=blue,breaklinks=true]{hyperref}

\long\def\symbolfootnote[#1]#2{\begingroup%
\def\thefootnote{\fnsymbol{footnote}}\footnote[#1]{#2}\endgroup} 

\begin{document}
 
\title{Comparing air quality statistical models}
\date{}
\author{Michela Cameletti\textsuperscript{a}, Rosaria Ignaccolo\textsuperscript{b}, \\Stefano Bande\textsuperscript{c}}

\maketitle

\vspace*{-1cm}
\begin{center}
\begin{small}
\textsuperscript{a}Dip. di Matematica, Statistica, Informatica e Applicazioni, \\ Universit\`a di Bergamo\symbolfootnote[1]{Corresponding address: Via dei Caniana 2, 24127 Bergamo (Italy). \\ E-mail: michela.cameletti@unibg.it}\\
\textsuperscript{b}Dip. di Statistica e Matematica Applicata, Universit\`a di Torino\symbolfootnote[2]{R. Ignaccolo is also affiliated with Statistics Initiative, Collegio Carlo Alberto, Italy.}\\
\textsuperscript{c}Dip. Tematico Sistemi Previsionali, Qualit\`a dell'aria, ARPA Piemonte\\
\end{small}
\end{center}

\begin{abstract}

Air pollution is a great concern because of its impact on human health and on the environment. Statistical models play an important role in improving knowledge of this complex spatio-temporal phenomenon and in supporting public agencies and policy makers. We focus on the class of hierarchical models that provides a flexible framework for incorporating spatio-temporal interactions at different hierarchical levels. The challenge is to choose a model that is satisfactory in terms of goodness of fit, interpretability, parsimoniousness, prediction capability and computational costs. In order to support this choice, we propose a comparison approach based on a set of criteria summarized in a table that can be easily communicated to non-statisticians. Our proposal - simple in principle but articulated in practice - holds true for many environmental phenomena where a hierarchical structure is suitable, a large-scale trend is included and a spatio-temporal covariance function has to be chosen.

We illustrate the details of our proposal through a case study concerning particulate matter concentrations in Piemonte region (Italy) during the cold season October 2005-March 2006. From the evaluation of the proposed criteria for our case study we draw some conclusions. First, a model with a complex hierarchical structure is globally preferable to one with a complex spatio-temporal covariance function. Moreover, in the absence of suitable computational resources, a model simple in structure and with a simple covariance function can be chosen, since it shows good prediction performance at reasonable computational costs.

\noindent \textbf{Keywords}: Particulate Matter PM$_{10}$, hierarchical models, spatial mapping, spatio-temporal covariance function, prediction performance indexes.

\end{abstract}
%
%

\section{Introduction}\label{sec:intro}

Air quality is jeopardized by the presence of several pollutants. Particulate matter (PM) is one of the most critical air pollutants in Europe and, despite the improvements thanks to European Union legislation, it still has a heavy toll on human life \citep{whynotfalling}. 
From a statistical perspective, many articles have been proposed for modelling the concentration of PM (and of other pollutants) and understanding its underlying complex spatio-temporal dynamics. 
In particular, almost all the works propose spatial prediction techniques in order to obtain concentration maps useful for evaluating the health risk and assessing compliance with European and national directives, even in places where no measurement stations are located (e.g. \citealp{GDC, GRASPA31_publ, STfinePM, sahunicolis, zidek}). Moreover, some works also develop methods for temporal forecasting (such as, for example, \citealp{bayesiankkf} and \citealp{pm2.5smith}) or consider more critical pollutants at the same time (e.g. \citealp{kuwait, shaddick, GRASPA32_publ}).  
When explanatory variables are available, measured by a monitoring network or simulated by a deterministic model, they contribute to the mean structure of the model, also known as large-scale component in geostatistical literature (see, for example, \citealp{cressie93}). To this regard sensitivity analysis techniques can be used to understand the role of covariates with respect to the output uncertainty, as shown in \cite{cocchi}  and \cite{simulation08}.

The common characteristic of many models found in literature is their hierarchical structure. This means that they are constructed by putting together conditional sub-models defined at each hierarchical stage. With reference to likelihood, this corresponds to taking a conditional point of view for which the joint probability distribution of a spatio-temporal process can be expressed as the product of certain simpler conditional distributions. This property makes it possible to deal with the complexity of spatio-temporal processes in a straightforward way which is the reason why hierarchical models have become so popular for modelling environmental processes, especially from a Bayesian perspective \citep{why, hierarchical_spacetime, hierarchical}.

With particular reference to air pollution, covariates, such as meteorological and orographical variables, play an important role in seizing the large-scale variation of data, the influence of meteorology and geographical factors on pollutant concentration being well-known. Residual variability, also called small-scale variation, is modelled by a spatio-temporal process defined at a particular level of the hierarchy and by a spatio-temporal covariance function. In this context, an interesting question arises concerning the best hierarchical structure that can be combined with the spatio-temporal covariance function in order to seize the spatial and temporal dynamics of the considered process. For example, is it preferable to have a two-level model with a complex nonseparable covariance function or a model with a more complicated hierarchical structure but a simpler covariance function? Finding an answer to this question might be useful in order to provide environmental agencies with an effective statistical model for building reliable PM concentration maps, equipped with the corresponding uncertainty measure.

The goal of this paper is to provide an instrument to answer to this question. Therefore, we propose certain criteria that are as objective as possible for comparing spatio-temporal models for air quality data. Such criteria consider both goodness of fit and model complexity as well as computational costs. 
Actually, a model comparison for space-time models is proposed by \cite{huangcomparison} through  AIC and BIC criteria, which are investigated in their practical behaviour since asymptotic theory of these indexes for geostatistical models is substantially missing in literature. Moreover, these authors use mean squared prediction error at a fixed time to compare prediction capability. Specifically for fine particulate matter,  \cite{pangcomparison} compare, on practical grounds, ordinary kriging with Bayesian maximum entropy technique, as implemented in SEKS-GUI software through averaged estimation errors and error variances at four validation sites \citep{SEKSGUI}.

Hence, in the absence of a space-time model selection theory, we first discuss the hierarchical models considered and then propose a set of empirical criteria to compare their intrinsic and computational complexity and their spatial prediction capability (through suitable indexes for air quality models opportunely summarized in a qualitative index). To implement our proposal, we deal with a case study concerning particulate matter in the Piemonte region (Italy). Here, we compare models on air quality real data; nonetheless, the proposed comparison approach holds true for general environmental phenomena where a hierarchical structure is suitable, a large-scale trend is included and a spatio-temporal covariance function has to be chosen. 

The paper is organized as follows. In Section \ref{sec:models} we introduce six models to be compared, specifying their hierarchical structure and spatio-temporal covariance function. The models, which constitute an extensive class of spatio-temporal hierarchical models, are fully discussed in Section \ref{sec:discussion}. Section \ref{sec:comparison} describes the criteria used to establish which model best describes the data. In particular, we consider the intrinsic and computational complexity as well as the spatial prediction capability of each model. The paper also features an application proposed in Section \ref{sec:application} regarding PM data measured in Piemonte during the 2005-2006 winter season.  Unlike \cite{huangcomparison} we implement all the models in a fully Bayesian framework via Markov Chains Monte Carlo (MCMC) methods. The paper ends with a discussion of the results and the conclusions. The Appendices contain details about the full conditional and predictive distributions involved in the model estimation and spatial prediction procedures which may help.

\section{Hierarchical spatio-temporal models}\label{sec:models}
Let $z\left(s_i,t\right)$ be the scalar spatio-temporal process observed at site $s_i$ and at time $t$ where $i=1,\ldots,d$ and $t=1,\ldots,T$.\\
We assume the following \textit{measurement equation} as the first level of the hierarchical models:
\begin{equation}\label{meas.eq}
z\left(s_i,t\right)=u\left(s_i,t\right)+\varepsilon\left(s_i,t\right)
\end{equation}
where $\varepsilon\left(s_i,t\right)\sim N\left(0,\sigma^2_\varepsilon\right)$ is the measurement error defined by a Gaussian white-noise process, serially and spatially uncorrelated. The term $u\left(s_i,t\right)$ is the so-called state process and can be defined, in turn, by other sub-levels giving rise to different hierarchical models described in the following subsections.

\subsection{Model A} \label{modelA}
Model A is a two-level regression model characterized by a large-scale term and a spatio-temporal process for the residual small-scale component. In particular, quantity $u\left(s_i,t\right)$ of Eq.\eqref{meas.eq} is given by the sum of a trend $\mu\left(s_i,t\right)$ and a random process $\omega\left(s_i,t\right)$, as follows:
\begin{equation}\label{modelAeq2}
u\left(s_i,t\right)=\mu\left(s_i,t\right)+\omega\left(s_i,t\right).
\end{equation}
Trend $\mu\left(s_i,t\right)$ is a function of $k$ covariates and is given by 
\begin{equation}\label{trend}
 \mu\left(s_i,t\right)=X\left(s_i,t\right)\beta
 \end{equation}
 where $X\left(s_i,t\right)=\left(X_{1}\left(s_i,t\right),\ldots,X_{k}\left(s_i,t\right)\right)$ denotes the covariate vector for site $s_i$ at time $t$ and $\beta=\left(\beta_{1},\ldots,\beta_{k}\right)^{\prime}$ is the coefficient vector.
 

The zero-mean Gaussian process $\omega\left(s_i,t\right)$ of Eq.\eqref{modelAeq2} is the residual process whose spatio-temporal covariance function depends on the parameter vector $\theta$, namely
\begin{equation}\label{STcovariance}
Cov\left(\omega\left(s_i,t\right),\omega\left(s_j,t^{\prime}\right)\right)=\sigma^{2}_{\omega}C_\theta\left(h,l\right) \qquad \forall i \neq j,t\neq t^{\prime}
\end{equation}
where $\sigma^{2}_{\omega}$ is the constant in time and space variance of the process and $C_{\theta}(.,.)$ is the spatio-temporal correlation function parameterized by $\theta$, with $h=\left\|s_i-s_j\right\|$ the Euclidean distance between sites $i$ and $j$, $l=\left|t-t^\prime\right|$ the temporal lag between $t$ and $t^\prime$. As the covariance function \eqref{STcovariance} only depends upon $h$ and $l$, the process is supposed to be second-order stationary and spatially isotropic. We consider three different forms for the covariance function, which give rise to the following models.
\begin{description}
\item[Model A1] Under the hypothesis that the process $\omega\left(s_i,t\right)$ is i.i.d. over time, it holds that
\begin{equation}\label{iidCov}
Cov\left(\omega\left(s_i,t\right),\omega\left(s_j,t^{\prime}\right)\right) =\left\{
\begin{array}
[c]{ccc}%
0 &  &  \mathit{if} \qquad t\neq t^{\prime}\\
\sigma^2_\omega C_\theta(h) && \mathit{if} \qquad t=t^\prime
\end{array}
\right. 
\end{equation}
where $C_\theta(h)$ is a purely spatial correlation function. To this regard many functions defining isotropic second-order stationary spatial processes can be found in \citet[Ch.1]{hierarchicalmodeling}.
\item[Model A2] Adopting a separable approach, the space-time covariance function factors into a purely spatial and a purely temporal component and is given by 
\begin{equation}
C_\theta\left(h,l\right)=C_{\theta_{1}}\left(l\right)C_{\theta_{2}}\left(h\right)\label{eq:sepCov}
\end{equation}
with $\theta=\left(\theta_{1},\theta_{2}\right)$. In this case there is no interaction between space and time.

\item[Model A3] Adopting a nonseparable approach, a space-time correlation function can be defined in a general form, as introduced in \cite{Gneiting}: 
\begin{equation}\label{covGNE}
C_\theta\left(h,l\right)=  \frac{1}{\psi(|l|^2)} \; \varphi\left( \frac{h^2}{\psi(|l|^2)}\right).
\end{equation}
The function $\psi(x), x\geq0$, is any completely monotone function while $\varphi(x), x\geq 0$, is any positive function with a completely monotone derivative. 
In particular,  we consider the two cases described in Table \ref{Tab:nonseparable}, corresponding to  {\bf Model A3-1} and {\bf Model A3-2} respectively, where $0 < b \le 1$, $\nu>0$, the smoothness parameters $\alpha$ and $\gamma$ take values in $(0, 1]$ and the space-time interaction parameter $\tau$ of Model A3-2 is defined in $[0, 1]$. Finally, the scaling parameters of time and space $a$ and $c$ are positive. It follows that $\theta_{A3-1}=\left(\alpha,b,\gamma, a, c \right)$ for Model A3-1 and  $\theta_{A3-2}=\left(\alpha,\tau,\gamma,\nu, a, c \right)$ for Model A3-2.
\end{description}
\subsection{Model B} \label{modelB}
Differently from Model A, Model B has a three-level hierarchical structure due to an additional equation required for modeling the temporal dynamics of a constant in space latent process. This means that for Model B the term $u\left(s_i,t\right)$ of Eq.\eqref{meas.eq} is given by the sum of a trend $\mu\left(s_i,t\right)$, of a unidimensional latent process $y\left(t\right)$ and of a purely spatial process $\omega\left(s_i\right)$, as follows:
\begin{equation}\label{modelBeq2}
u\left(s_i,t\right)=\mu\left(s_i,t\right)+y\left(t\right)+\omega\left(s_i\right),
\end{equation}
where $y(t)$ and $\omega\left(s_i\right)$ are uncorrelated. In Eq.\eqref{modelBeq2} the trend $\mu\left(s_i,t\right)$ is defined as in Eq.\eqref{trend}, while the term $y\left(t\right)$ refers to a constant in space unidimensional process with a temporal dynamics, with an autoregressive coefficient $\rho$,  given by
\begin{equation}\label{modelBeq3}
y\left(t\right)=\rho y \left(t-1\right)+\eta \left(t\right)
\end{equation}
where $y\left(0\right)\sim N\left(0,\sigma^{2}_{B}\right)$ and $\eta\left(t\right)\sim N\left(0,\sigma^{2}_{\eta}\right)$ are uncorrelated. Finally the spatial process $\omega\left(s_{i}\right)$ is assumed to follow a Gaussian distribution $N\left(0,\sigma^{2}_{\omega}\right)$ whose spatio-temporal covariance function is
 $Cov\left(\omega\left(s_i\right),\omega\left(s_j\right)\right)=\sigma^{2}_{\omega}C_\theta\left(h\right)$. 
 
\subsection{Model C}
Model C is a three-level hierarchical model defined by a spatio-temporal residual process with temporal dynamics. Specifically the term $u\left(s_i,t\right)$ of Eq.\eqref{meas.eq} is given by a trend $\mu\left(s_i,t\right)$ and a spatio-temporal process $y\left(s_i,t\right)$, namely
\begin{equation}\label{modelCeq2}
u\left(s_i,t\right)=\mu\left(s_i,t\right)+y\left(s_i,t\right).
\end{equation}
The trend $\mu\left(s_i,t\right)$ has the same structure as in Model A and B (see Eq.\eqref{trend}) while $y\left(s_i,t\right)$ is a spatio-temporal process that changes in time according to the following dynamics with autoregressive coefficient $\rho$:
\begin{equation}\label{modelCeq3}
y\left(s_i,t\right)=\rho y\left(s_i,t-1\right)+\omega\left(s_i,t\right)
\end{equation}
where $y\left(s_i,0\right)\sim N\left(0,\sigma^{2}_{C}\right)$. Finally, $\omega\left(s_i,t\right)\sim N\left(0,\sigma^{2}_{\omega}\right)$ is a spatio-temporal process i.i.d. over time, so that its spatio-temporal covariance function is given by Eq.\eqref{iidCov}. Obviously $y$ and $\omega$ are uncorrelated.

\section{Model discussion}\label{sec:discussion}
All the models are characterized by the same measurement equation given by \eqref{meas.eq} and by the trend term in Eq.\eqref{trend} defined as a function of some covariates that can change in space and time (e.g. meteorological variables) or can be constant in time (e.g. spatial coordinates).  The six models we consider differ in the way the residual detrended process is modelled and in how the spatio-temporal correlation is treated. 

Model A, described in Section \ref{modelA}, features a very simple structure characterized by a unique residual spatio-temporal process for which three different covariance functions are considered. The most complex case is represented by Model A3, which is characterized by a nonseparable spatio-temporal covariance function. Considering that space and time interact together, from a physical point of view it is reasonable to adopt a nonseparable approach even if it has certain computing drawbacks. In fact, the size of the variance-covariance matrix of the $\omega\left(s_i,t\right)$ process is $(dT \times dT)$ and, especially in the case of a dense network or a long monitoring period, matrix operations become infeasible from a computational point of view.  Model A1 and Model A2 introduce some simplifications: Model A1 supposes that the process $\omega\left(s_i,t\right)$ is i.i.d. over time, which leads to a purely spatial covariance function (see Eq.\eqref{iidCov}), while Model A2 is based on the separability hypothesis (see Eq.\eqref{eq:sepCov}). The choice of a spatio-temporal covariance function to be used depends on certain considerations. First of all, as stated in \cite{huangcressie}, separable models are often chosen for convenience rather than for their ability to properly fit the data; basically, the same holds true for the i.i.d. over time hypothesis case. This is mainly related to the computational advantages in implementing and estimating a model with these simplified functions, which usually depend on a small number of parameters and involve smaller matrices. Generally speaking, a spatial covariance function that does not depend on time - the i.i.d. case - can be used when it is possible to show, for example by means of daily variograms, that the spatial correlation does not change significantly in time. If this is not the case, a separability test (e.g. \citealp{fuentesseparability}) should be performed in order to verify if the separability hypothesis can be assumed. Otherwise, a nonseparable covariance function should be used.

Model B and C differ from Model A for their three-level structure: in both cases, in fact, an equation is introduced for modelling the temporal dynamics of a latent process. In particular, for Model B this is given by a purely temporal AR(1) process (defined in \eqref{modelBeq3}) which is supposed to be constant in space, while for Model C it has an AR(1) structure  with innovations i.i.d. over time (see Eq.\eqref{modelCeq3}).
It is important to point out that the size of the latent process is always unidimensional for Model B while for Model C it is defined by the number of spatial sites $d$. From a computational perspective this means that estimation procedure costs are higher for Model C than for Model B, even if Model B has an extra parameter to estimate, i.e. $\sigma^2_\eta$.  

It is interesting to point out that both the spatio-temporal covariance functions of Model B and Model C can be rewritten in a separable form, additive and multiplicative respectively (for further details see Appendix \ref{app0}). 

\section{Model comparison}\label{sec:comparison}
We are interested in determining which is the most effective model for fitting the data. To achieve this goal, we compare the six models using a set of empirical criteria which explore the model complexity and prediction capability, as described hereafter.

\subsection{Intrinsic and computational model complexity}
The intrinsic complexity of a model can be roughly defined as the number of parameters to be estimated. Generally, as the number of parameters increases the estimation procedure becomes more complex since, in the fully Bayesian framework we adopt, steps are added to the Gibbs sampling algorithm. Moreover, if the parameter estimation requires the use of the Metropolis-Hastings (MH) algorithm, since no closed-form full conditional posterior distributions are available, the algorithm can become unstable and requires a larger number of iterations in order to reach convergence. This underscores out how a richer model, in terms of parameters and hierarchical structure, is necessarily more complex from a computational point of view. To this regard, we also consider the size of the biggest matrix to be inverted for each model. Considering that matrix inversion is of order $n^3$ in computation (where $n$ is the total number of data), a considerable size can give rise to massive computational loads infeasible to be carried out. To this regard, just to have an idea, consider that Matlab\footnote{We use Matlab R2009b with the Parallel Computing Toolbox.} on an Intel Core 2 Duo Mac (2.4 Ghz, 4GB RAM) takes about 15 and 76 seconds to invert - using the Cholesky factorization - a full symmetric matrix with size $(4500\times 4500)$ and $(8000\times 8000)$, respectively. The same operations require about 5 and 21 seconds on an Intel Xeon 8 CPU cluster (2.66 Ghz,  8 GB RAM). In the hypothetical case of one parameter and 100000 iterations required for the convergence of the algorithm, at least 5 or 24 days would be necessary for the implementation of the 4500 or 8000-dimensional cases, respectively. Obviously, these computing times increase (in a non linear way) as the parameter set becomes larger. 

This information about intrinsic and computational model complexity is shown in the first rows of Table \ref{Tab:comparison} (pag. \pageref{Tab:comparison})\footnote{We suppose that for Model A1, B and C the parameter vector $\theta$ is unidimensional and that for Model A2 the parameter vectors $\theta_1$ and $\theta_2$ are unidimensional.}. It is clear that the nonseparable models, Model A3-1 and Model A3-2, are the most complex ones since they have the biggest parameter vectors (respectively 7 and 8 parameters, excluding the $\beta$'s), all estimated using the MH algorithm. Moreover, the size of the variance-covariance matrix is $(dT\times dT)$. This means that, from a computational point of view, nonseparable models are extremely expensive and their implementation is expected to be severely time consuming.

Model A2 has 4 parameters to be estimated and, thanks to the separability of the spatio-temporal covariance function, it enjoys the properties of the Kronecker product. This results in certain computational advantages regarding the inverse and the determinant of the $(dT\times dT)$ variance-covariance matrix, because we deal with $(d\times d)$ and $(T\times T)$ matrices separately (for details see Section \ref{sec:A2}). In Table \ref{Tab:comparison} the size of the biggest matrix to be inverted is $(T\times T)$ since we consider a small monitoring network, such that  $d\ll T$.


In terms of complexity Model A1, Model B and Model C are more suitable because they have small parameter vector, make use of the MH algorithm in a limited way and are characterized by $(d\times d$)-dimensional variance-covariance matrix. This means much more computationally manageable models. 

With respect to the computational complexity, the models are compared also considering the computing time required to estimate the parameters and for performing the spatial predictions over the validation stations. The times are evaluated per iteration of the MCMC run and are computed using the above-defined Intel Xeon 8 CPU cluster. Generally, it is clear that, keeping all the rest equal, a model that can be quickly implemented is more desirable.


\subsection{Spatial prediction capability}\label{sec:indexes}
As the aim of the modelling is prediction, we compare models on the basis of their spatial prediction capability which is evaluated using certain performance indexes computed on validation stations.
In particular, we consider five indicators based on the differences between predicted and observed data together; moreover, we compute the observed coverage probability. More precisely, together with the usual root mean square error (RMSE) and the correlation coefficient $\rho$, we adopt the Normalised Mean Bias Factor (NMFB) recently introduced by \cite{Yu} and two indexes, named WNNR and NNR, proposed by \cite{Poli} and defined on the Normalized Ratios between the predicted and the observed values.

For a fixed site $s_i$, let $z_t$ be the observed time series and $\hat z_t$ the predicted time series with $t=1,\ldots, T$ (see Appendix \ref{app2} for details about prediction); moreover, denote with $\bar{z}$ and $\bar{\hat z}$ the corresponding mean values. The Normalised Mean Bias Factor is defined by
\[
\text{NMBF} =
\left\{\begin{array}{ll}
    \frac{\sum_t \hat z_t}{\sum_t z_t} - 1 & \qquad  \text{if} \; \bar{\hat z} \ge \bar{z}\\
    1 - \frac{\sum_t z_t}{\sum_t \hat z_t} & \qquad \text{if} \; \bar{\hat z} < \bar{z}\\
    \end{array}\right..
\]
NMBF is defined on $\mathbb{R}$ and has the advantage of both avoiding inflation and asymmetry, two problems discussed in \cite{Yu}. 

The Weighted Normalised mean square error of the Normalised Ratios is defined by
\[
\text{WNNR}= \frac{\sum_t s_t^2 (1-k_t)^2}{\sum_t s_t k_t},
\]
while the non-weighted one is
\[
\text{NNR} = \frac{\sum_t (1-k_t)^2}{\sum_t k_t},
\]
where $s_t = z_t/ \bar{z}$ is the weight and $k_t= \exp{-|\ln(\hat z_t/z_t)|}$ is the normalised ratio. WNNR and NNR are both positive and have the advantage of taking properly into account the peaks of observed data (see the discussion in \citealp{Poli}).

These 5 indexes and the observed coverage probability are computed for all the validation stations and for each model. Successively, these 6 performance measures are summarized over stations in a qualitative index based on ``stars category"\hspace{-0.06cm}, so that more stars correspond to a better prediction capability of the model (see Table \ref{Tab:comparison} on page \pageref{Tab:comparison}).

\section{Model comparison for PM$_{10}$ in Piemonte}\label{sec:application}
In order to compare the models described in Section \ref{sec:models}, below we consider particulate matter concentration with an aerodynamic diameter of less than 10 $\mu m$ (PM$_{10}$, in $\mu g$/$m^{3}$) measured in the Piemonte region (Italy) during the October 2005 - March 2006 winter season. Piemonte is situated in the western part of the Po river basin, surrounded on three sides by the Alps. The pollutant dispersion is strongly influenced concurrently by the shelter effect of the Alps and by meteorological features which depend on the complex orography of the region. So, for example, we can have weak winds and stagnation in the central part of the region or breezes and foehn winds in mountains and valleys. For these reasons, we usually observe lower PM$_{10}$ concentration near the Alps, whereas higher pollution levels are detected in plains closer to urban areas. Actually, as expected, air quality is worse in urbanized areas where the most important emission sources, in other words industrial sites and high traffic levels, are located.

\subsection{Data description}
\begin{figure}[h]
\begin{center}
\includegraphics[scale=0.3]{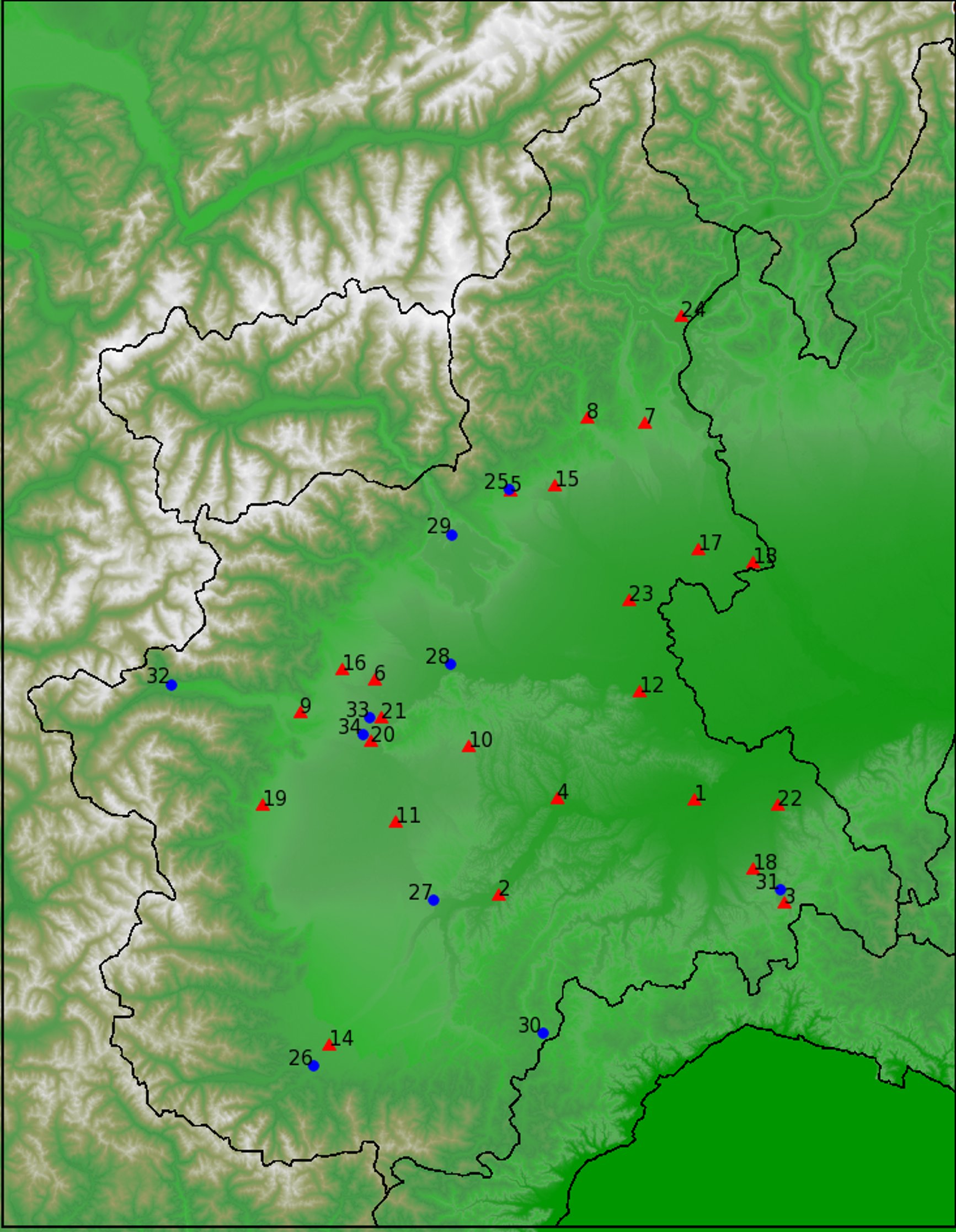}
\caption{Locations of the 24 PM$_{10}$ monitoring sites (red triangles) and 10 validation stations (blue dots). The complete names of the 24 stations are given in the x-axis labels of Fig.\ref{boxplot_urban}. The labels of the 10 validation sites are: 25 Biella - Largo Lamarmora, 26 Borgo San Dalmazzo, 27 Bra, 28 Chivasso, 29 Ivrea, 30 Saliceto, 31 Serravalle Scrivia, 32 Susa, 33 Torino - P.zza Rivoli, 34 Torino - Via Gaidano. }
\label{map}
\end{center}
\end{figure}

\begin{figure}[h]
\begin{center}
 \includegraphics[scale=0.45]{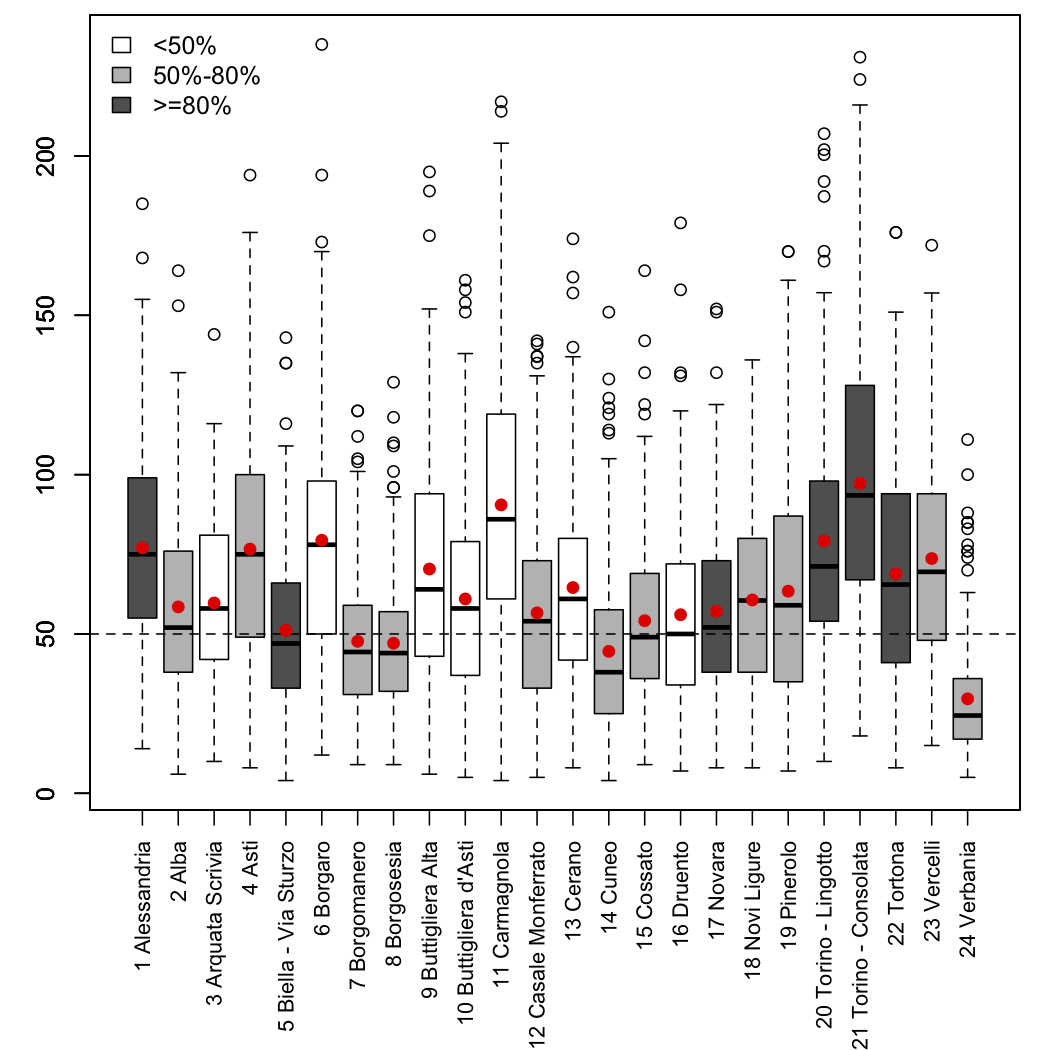} 
\caption{PM$_{10}$ concentration distribution over stations (the bold horizontal line corresponds to the median while the bold dot to the mean). The colours of the boxplots are given by the area type, i.e. the percentage of built-up surface (see the legend). }
\label{boxplot_urban}
\end{center}
\end{figure}

We analyze daily PM$_{10}$ data measured by a network of $d=24$ sites (see red triangles in Figure \ref{map} and the corresponding labels in Figure \ref{boxplot_urban}) for $T=182$ days (data are provided by an information system called \textit{AriaWeb Regione Piemonte}). Moreover, we set aside data from 10 sites for validation purposes (see blue dots in Figure \ref{map}). The 24 sites are selected so that the amount of missing data does not exceed 20\% and the missing data is not sequential. This guarantees good spatial coverage of the monitoring network so that stations can also be found in rural plain areas as well as in urbanized alpine valleys. 

Figure \ref{boxplot_urban} shows the distribution of the PM$_{10}$ concentration by stations. We can see that most of the stations have average and median concentration levels above 50 $\mu g$/$m^{3}$, which is the threshold set by the European Commission (2008/50/EC directive) that can be exceeded no more than 30 days a year.  All the distributions are positive skewed due to the occurrence of extremely polluted days. This kind of situation is quite common in the Po Valley, especially during the winter season when relatively stable atmospheric conditions, associated with a reduced washout of particulate matter, give rise to higher concentration.

According to European legislation, each site is classified by area-type which has three categories (rural, suburban and urban) depending on the level of urbanisation of the area. More precisely, in Piemonte a site is classified rural if the percentage of built-up surface within a 1 km radius around it is less than 50\%, suburban if that percentage is between 50\% and 80\%, urban if otherwise. Looking at Figure \ref{boxplot_urban}, area-type does not seem to have a visible effect on  PM$_{10}$ concentration. In fact, we observe  rural stations with high levels of pollution as well as urban locations with lower concentration. This can seem unusual since area-type can be considered as a proxy of anthropogenic activities, but it could happen since PM$_{10}$ pollution is a complex phenomenon strongly related to meteorology and orography, especially during the winter season.

Indeed, our models share a common trend component where some meteorological and orographical variables appear. The first ones are time-varying covariates obtained from a nested system of deterministic computer-based
models implemented by the environmental agency ARPA Piemonte \citep{Bande,deterministicmodels}. Such models provide the estimates on a regular 4 \textit{km} $\times$ 4 \textit{km} grid of some meteorological variables, turbulence and chemicals parameters. By means of a preliminary regression analysis using Akaike's information criteria (AIC) and parameter significance, we choose the following covariates: daily maximum mixing height (HMIX, in $m$),  daily total precipitation (PREC, in $mm$), daily mean  wind speed (WS, $m/s$), daily mean temperature (TEMP, in $K$) and daily emission rates of primary aerosols (EMI, in $g/s$). The mixing height is one of the fundamental parameters that characterize the structure of the atmosphere near the ground. Low mixing heights mean that the air is generally stagnant with very little vertical motion, and therefore pollutants are usually trapped near the ground surface. High mixing heights allow vertical mixing within a deep layer of the atmosphere and good dispersion of pollutants.  Thus a negative relationship is expected between PM$_{10}$ and HMIX. Precipitation plays a crucial role in explaining PM$_{10}$ variations taking  into account the wet removal of aerosol suspended in the atmosphere. Moreover, the daily mean values of wind speed are used in order to take into account the pollutant removal due to strong wind episodes, often combined with foehn conditions, that frequently occur in Piemonte during the winter time. Mean temperature influences the dispersion and accumulation of pollutants. On the one hand, it is related to atmospheric photochemical reactions and, consequently, to the production of secondary aerosols. On the other, low temperatures near the ground are often related to strong thermal inversion, one of the atmospheric features responsible for heavy pollutant events in urban area. Moreover, it is well-known that low temperatures cause an increase in particulate emissions from vehicle traffic sources. Finally, emissions take into account information about anthropogenic activities (e.g. energy production, domestic and industry production, road transport) which are the main sources of primary pollutants and the precursors of secondary pollutants. Together with these time-varying covariates, we also consider altitude (A, in $m$) and coordinates (UTMX and UTMY, in $km$) in order to take into account the orography of the region.

\begin{figure}
\begin{center}
 \includegraphics[scale=0.4]{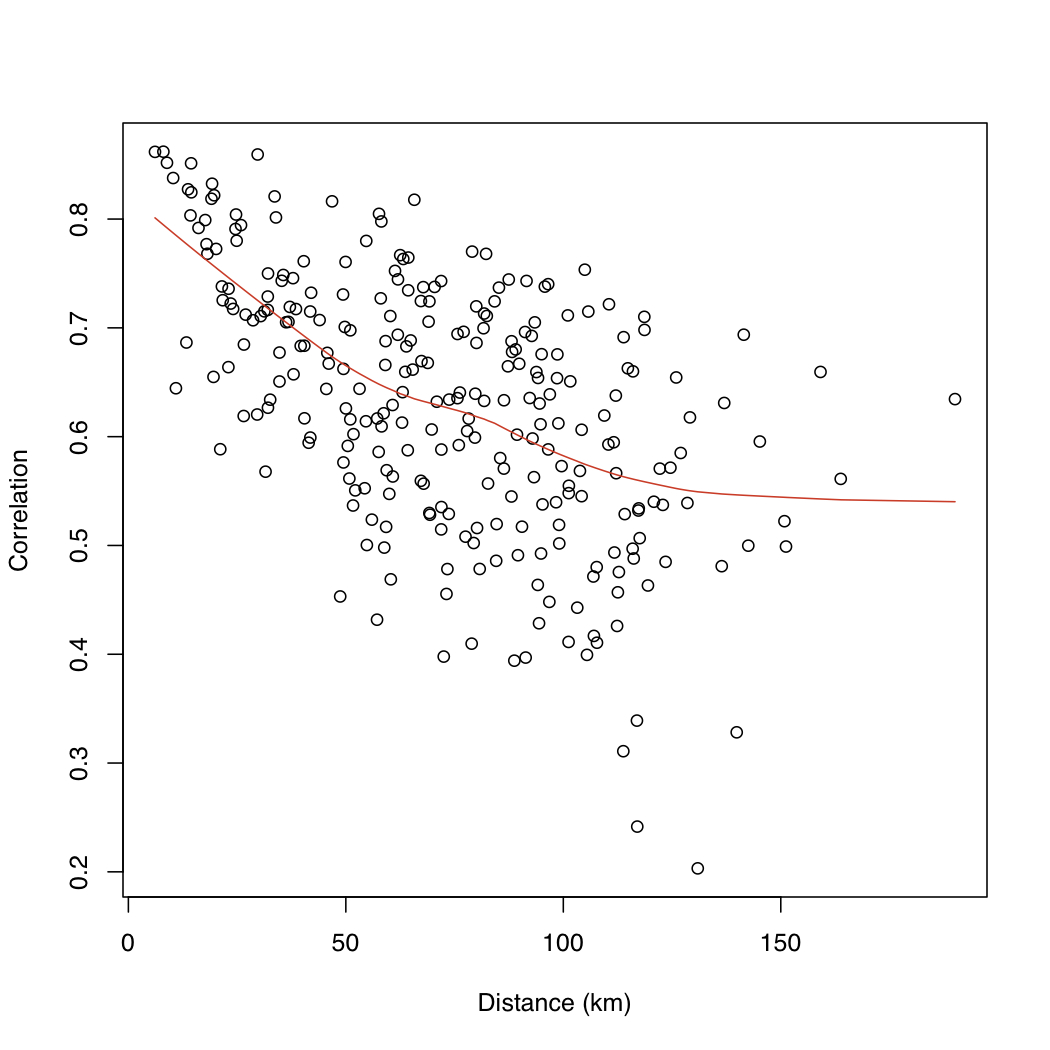} 
\caption{Spatial residual correlation cloud and \textit{lowess} curve.}
\label{fig:spatial_corr}
\end{center}
\end{figure}

\begin{figure}
\begin{center}
 \includegraphics[scale=0.4]{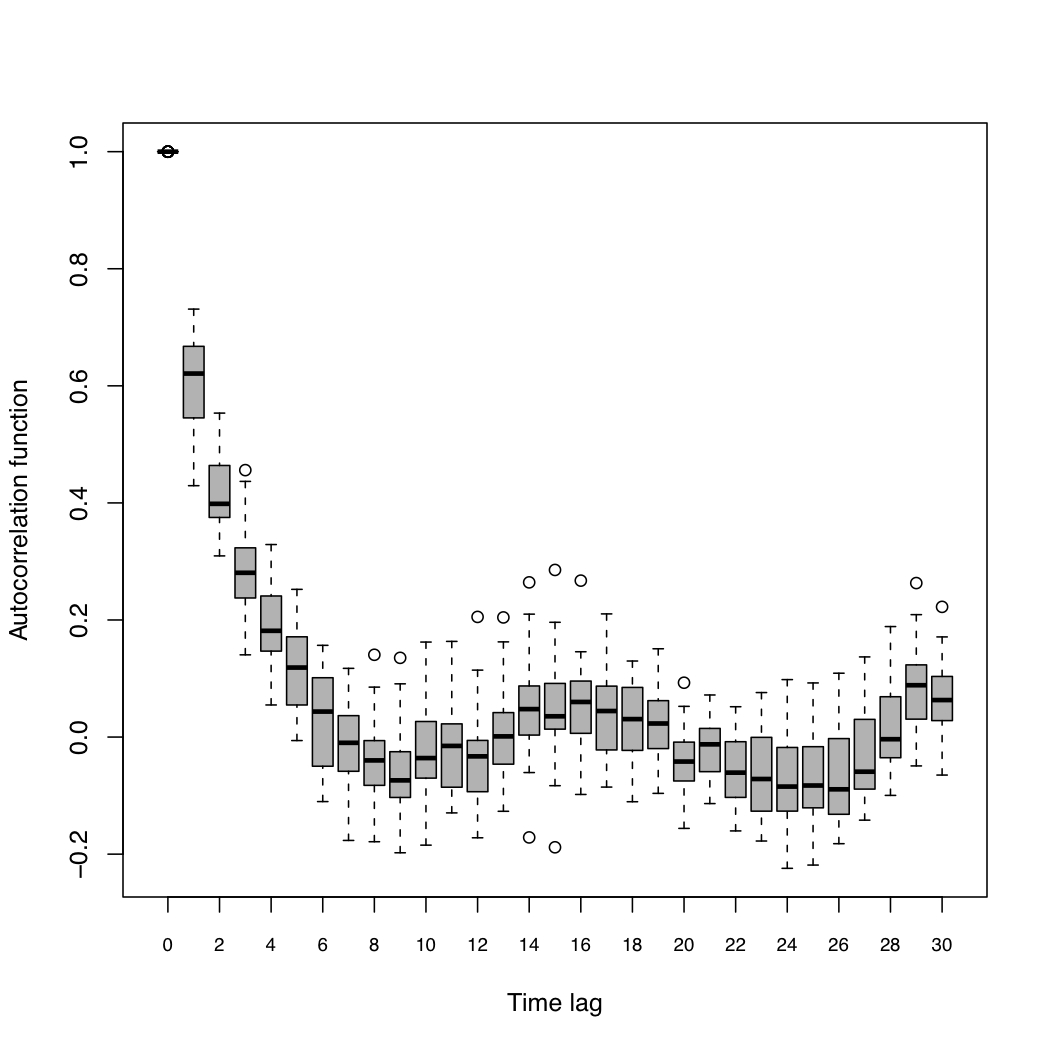} 
\caption{Boxplots of the residual empirical autocorrelations computed over the 24 stations.}
\label{fig:acf}
\end{center}
\end{figure}

In order to stabilize the variances, which increase with the mean values, and make the distribution of PM$_{10}$ data approximately normal, we adopt the logarithm transformation of PM$_{10}$ data. Subsequently, in order to investigate spatial and temporal correlation, we fit a simple regression model over all the 4368 (=$24\times 182$) PM$_{10}$ data with the covariates - standardized - described above. Figure \ref{fig:spatial_corr} illustrates the residual correlation cloud, that is the set of spatial correlations between sites at different distances. It shows that, even after removing the so-called large-scale component given by the covariates, a spatial correlation still remains: the lowess curve decreases slowly with distance and settles around 0.6 at 100 km. Figure \ref{fig:acf} shows the boxplots of empirical autocorrelations calculated on the residuals  - over stations - and it stands out that a temporal structure also remains, with a temporal correlation of about 0.6 at the first time lag. These results show great evidence of the importance of using a spatio-temporal model for catching the complex structure and dynamics of the phenomenon.

\subsection{Model implementation details}\label{sec:modelingdetails}
The models presented in Section \ref{sec:models} are now fitted on the PM$_{10}$ concentration data described in the previous section.

The covariate vector $X\left(s_{i},t\right)$ of Eq.\eqref{trend}, which also includes a constant for the intercept term, is given by
\begin{eqnarray*}
X\left(s_{i},t\right)&=&\bigg(1,A\left(s_{i}\right),UTMX\left(s_{i}\right),UTMY\left(s_{i}\right),WS\left(s_{i},t\right),\bigg.\\
&&\bigg. HMIX\left(s_{i},t\right),TEMP\left(s_{i},t\right),PREC\left(s_{i},t\right),EMI\left(s_{i},t\right)\bigg)
\end{eqnarray*}
where $i=1,\ldots,24$ and $t=1,\ldots,182$.

With reference to the covariance functions, we adopt a purely spatial exponential form given by $C_{\theta}\left(h\right)=\exp\left(-\theta h\right)$ for Model A1, B and C and a double exponential structure for Model A2, $C_{\theta}\left(h, l\right)=\exp\left(-\theta_{1} l\right)\exp\left(-\theta_{2} h\right)$. For Model A3-1 and A3-2 we refer to the nonseparable functions defined in Table \ref{Tab:nonseparable}.

As regards inference, i.e. parameter estimation and spatial prediction, we adopt a fully Bayesian approach via Monte Carlo Markov Chain (MCMC) methods. In particular, we use the Metropolis-within-Gibbs algorithm implemented in Matlab by ad-hoc written code. The complete details of the full conditional and predictive distributions are given in Appendices \ref{app1} and \ref{app2}. Convergence is diagnosed by monitoring the mixing of the chains by means of traceplots together with autocorrelation and kernel density plots. 

With reference to the prior specification, we assume Normal independent priors $N(0, 100)$ for each component of the $\beta$ vector and Inverse Gamma distributions $IG(2, 1)$ for the variance parameters ($\sigma^2_\omega, \sigma^2_\varepsilon$ and $\sigma^2_\eta$). A Uniform prior distribution $U(0,1)$ is chosen for the parameter $\theta$ of Model A1, B and C, and the parameter $\theta_2$ of Model A2: this corresponds to a spatial correlation between 0.37 and 1 at 1 $km$ of distance and between 0 and 1 at the maximum distance of 190 $km$. For the temporal decay parameter $\theta_1$ of Model A2 we adopt a Uniform prior distribution $U(0.3,3)$ whose support corresponds to a temporal range between 1 and 10 days (the range is determined using the relationship $\exp(-\theta_1 l)\approx 0.05$). For the parameters of the nonseparable correlation function of Model A3-1 and A3-2, we suppose vague prior distributions, that is $U(0,1)$ for $\alpha, b, \gamma, \tau$ and $U(0,10)$ for $a, c, \gamma$. 

Note that for Model A2 it is not possible to marginalize the posterior distribution over the latent process $u(s_i,t)$ (see Appendix \ref{app1} for details); thus, for sampling it we employ the \emph{en bloc} procedure described in \cite{STfinePM}. Finally, for Model B and C the latent process $y$ is sampled from its full conditional distribution by the Forward Filtering Backward Sampling (FFBS) algorithm as described in \cite{FFBS} . 


\subsection{Results and discussion}\label{Sec:results}

Table \ref{Tab:beta} displays the posterior estimates for the covariate coefficient vector $\beta$. Note that they are robust with respect to the model specification, meaning that the choice of a spatio-temporal covariance function as well as the introduction of a temporal dynamics equation do not affect strongly the estimates. In particular, the posterior mean of the intercept is around 3.9 on the log scale, which corresponds to an average pollution level of about 49.4 $\mu g/m^3$. As expected, a significant and positive relationship can be seen between emissions (EMI) and  PM$_{10}$ concentration. Moreover, the significance of the coefficients of  WS, HMIX, TEMP and PREC confirms the importance of meteorological variables on air quality. The UTMX and UTMY coefficient estimates are both negative, indicating a decreasing spatial trend from  West to East and from South to North. Finally, altitude (A) has a significant effect in reducing PM$_{10}$ concentration. 

With regards to the variance estimates in Table \ref{Tab:variances}, we observe that more variation is explained by the spatial or spatio-temporal term rather than by the measurement error. For Model A1, A3-1, A3-2 and B, the estimate of  $\sigma^2_\omega$ is larger - between 1 and 3 times - than the estimate of $\sigma^2_\varepsilon$ while for Model A2 and C it is, respectively, 16 and 74 times larger. The fact that for Model C the measurement error variance is so small with respect to the spatio-temporal error variance can be ascribed to its complex structure, clearly useful for explaining variability of the spatio-temporal process.

Table \ref{Tab:correlation} reports the posterior estimates of the correlation function coefficients. For Model A1, the estimate of $\theta$ is 0.0033 giving rise to a strong spatial correlation which decreases slowly with distance. In fact, at a maximum distance of 190 $km$ the spatial correlation is about 0.54. For Model A2, we find $\theta_1=0.492$ and $\theta_2=0.032$ that correspond to a temporal and spatial range of, respectively, 6 days and 95 $km$ (the spatial range is determined using the relationship $\exp(-\theta_2 h)\approx 0.05$). Instead, for Model B we estimate a range of about 60 $km$, indicating that the spatial correlation decreases rapidly and that at 190 $km$ there is no correlation. On the other hand, for Model B the estimate of the AR(1) temporal correlation coefficient $\rho$ is 0.8313, confirming the short-term temporal persistence of particulate matter. Model C seems to have a spatial correlation similar to the one of Model A1 (the posterior mean of $\theta$ is 0.0022 with a correlation of about 0.66 at 190 $km$), while its temporal correlation is weaker (estimate of $\rho$ equal to 0.6535) than the one of Model B. 

\begin{figure}[h]
\begin{center}
\includegraphics[scale=0.35]{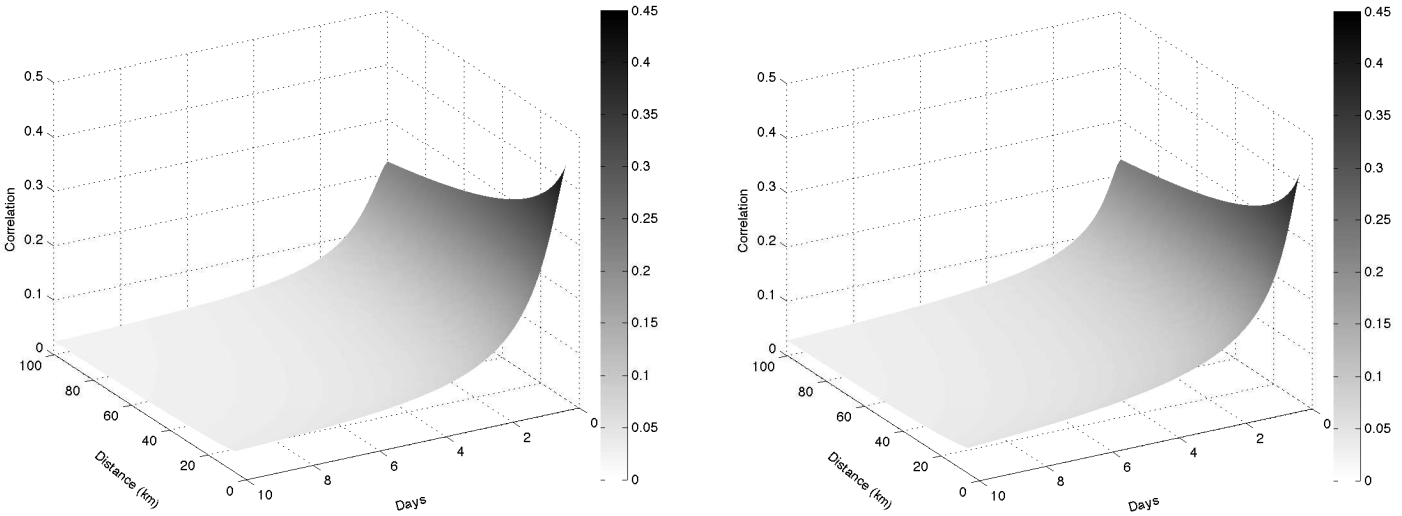}
\caption{Nonseparable spatio-temporal correlation function for Model A3-1 (left) and Model A3-2 (right) with the estimated parameters in Table \ref{Tab:nonseparable_par}.}
\label{ModA31_2corr}
\end{center}
\end{figure}

Posterior means and credible intervals of the nonseparable correlation function parameters (Model A3-1 and A3-2) are shown in Table \ref{Tab:nonseparable_par}. Using these estimates, the correlation functions defined in Eq.\eqref{covGNE} and Table \ref{Tab:nonseparable} are plotted in Figure \ref{ModA31_2corr} (with 0-10 days of temporal lag and 0-100 $km$ of spatial distance). The two surfaces resemble each other, in other words the spatio-temporal correlation decreases similarly with respect to both the temporal lag and the spatial distance.

As discussed in Section \ref{sec:discussion}, the first rows of Table \ref{Tab:comparison} provide information about intrinsic and computational model complexity. Moreover, the table contains estimation and prediction times (in seconds per iteration). As expected, the estimation computing time increases with the number of parameters and the size of the biggest matrix to be inverted: in particular, Model A3-1 and A3-2 are the most time consuming (about 45 and 47 seconds, respectively) while the simple Model A1 is the fastest to be estimated (0.014 seconds per iteration).  Focusing on Model B and C, both characterized by an additional equation for the temporal dynamics requiring the use of the FFBS algorithm (see \eqref{modelBeq3} and \eqref{modelCeq3}), we observe that Model B is less time-consuming than Model C, even if it has one additional parameter to estimate, $\sigma^2_\eta$, and adopts more extensively the MH algorithm. This can be explained by considering that, while for Model B the FFBS algorithm involves 1-dimensional terms, for Model C it deals with $d$-dimensional terms. 
Model A2 requires more time (about 4 seconds) than Model A1, B and C because it uses \emph{en bloc} sampling for $u(s_i,t)$, that is an iterative computationally expensive procedure.

Regarding prediction time, Model A1 and B have very similar performances since the prediction procedure is almost the same (see Appendix \ref{app2}). Model C is slightly slower because it has a more complicated temporal dynamics requiring the introduction of an additional step for the composition sampling procedure given by Eq.\eqref{eq:cond posterior C}. The prediction requires more time for Model A3-1 and A3-2 given that the routine involves variance-covariance matrices with size $(dT\times dT)$. The 27 seconds per iteration necessary for Model A2 could be a cause for worry, but they are justified by the fact that drawing from the Normal distribution, as done in Eq.\eqref{eq:cond posterior A2}, slows down the composition sampling procedure.

\begin{figure}[h]
\begin{center}
\includegraphics[scale=0.5]{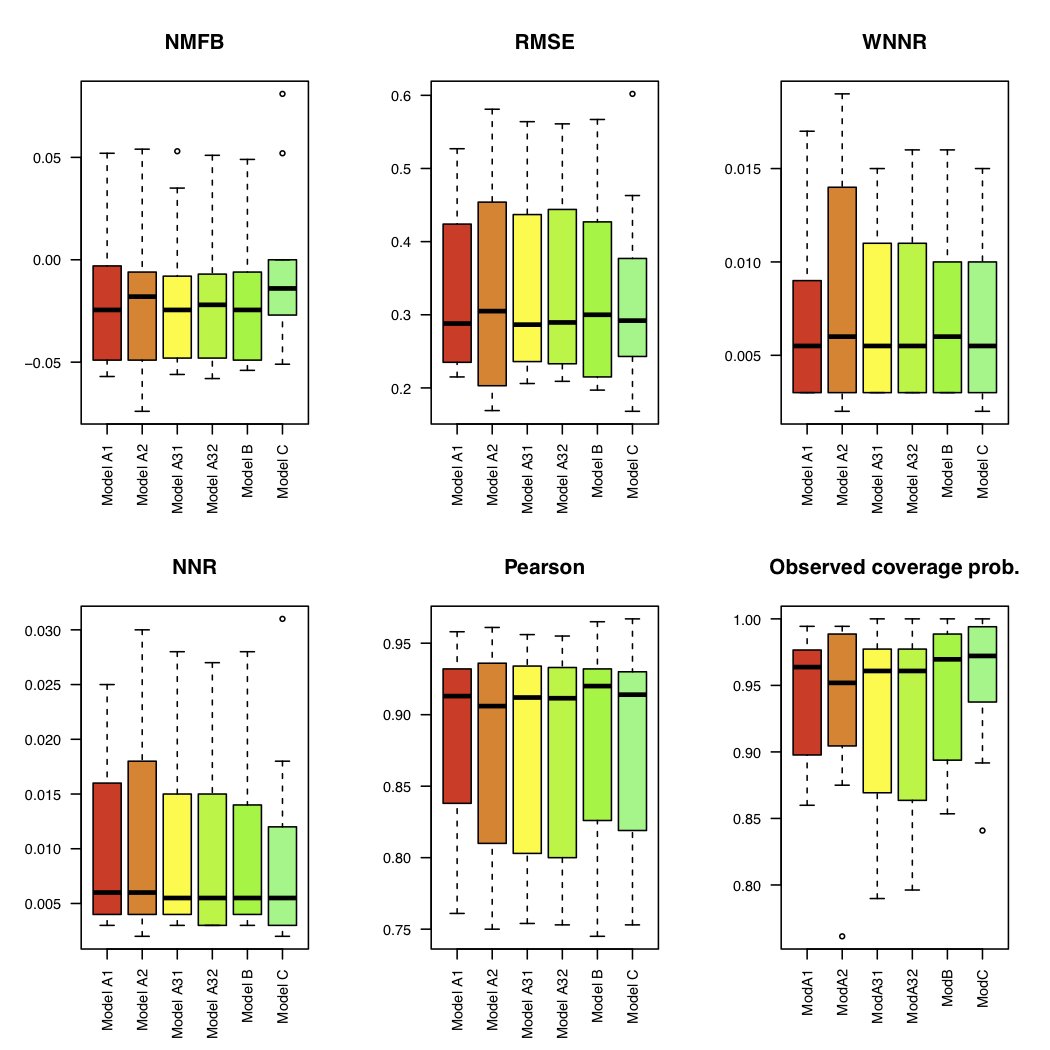}
\caption{Boxplots of the performance measure distributions computed for each model over the 10 validation stations.}
\label{boxplotindex}
\end{center}
\end{figure}

The last row of Table \ref{Tab:comparison} summarizes in ``stars category" (from 1 to 3) the prediction capability of our models. The stars are based on the values of the performance measures defined in Section \ref{sec:indexes} and computed for the 10 validation sites. Figure \ref{boxplotindex} shows through boxplots, for each model, the distribution of the indexes and of the observed coverage probability over station. 
The boxplots of Model A2 show the greatest variability and the largest values of RMSE, WNNR and NNR, meaning that the prediction is poorer where the observed time series have peaks. Thus, overall Model A2 has the worst prediction performance and it is classified with one star. Conversely, Model C boxplots show less variability, especially looking at NMFB, RMSE and the observed coverage probability. The other models seem to have very similar prediction performances, except for the observed coverage probability distributions of Model A3-1 and A3-2 that are left skewed (their first quartiles are around 0.87). For this reason we assign three stars to Model C and two stars to the remaining models.

The entire Table \ref{Tab:comparison} provides the set of criteria useful for choosing the most effective model in our case study. First of all, we discard Model A2 because of its poor prediction performance and long computing time for spatial prediction. Also, we can not suggest implementing Model A3-1 and A3-2 - with the complex nonseparable covariance function - for reasons of computational costs that are not offset by a significant improvement of the prediction capability. Among the remaining models, Model B is the less appealing one since it has less stars than Model C and is slower than Model A1 in the estimation routine. Finally, the choice is between Model A1 - simple in structure and covariance function - and Model C, which differs in the hierarchy for an additional equation explicating the temporal dynamics. As observed, Model C has a slightly better prediction capability, which is offset by a larger computational cost.  In the absence of suitable computational resources, Model A1 has a good performance at a reasonable cost.

\section{Conclusions}
In this work we propose a comparison of six models for air quality data, taking into account different levels of complexity either in the hierarchical structure  or in the spatio-temporal covariance function. This makes it possible to analyze a wide range of models suitable for complex environmental phenomena characterized by both a large and small scale of variation, as well as by spatial relationships, temporal dynamics and spatio-temporal interactions. In this work we focus on PM concentration, comparing models on the basis of certain criteria that take into account intrinsic complexity, computational costs and spatial prediction capability. These criteria are particularly important because environmental agencies need to evaluate air quality status, predicting pollutant concentration at unmonitored sites, at a reasonable computational cost. 
The table that summarizes the proposed set of criteria - Table \ref{Tab:comparison} in our case study - has the merit of being easy to communicate to non-statisticians, who may be environmental agency practitioners or policy makers, responsible for drawing up air quality legislations. Moreover, these spatial prediction maps can be used in risk assessment analysis and ecological risk models, as continuous exposure levels, or as validation criteria in a network design study. 

The compared models share a large-scale trend component whose estimated coefficient are robust as the model changes. Furthermore, this property allows models to be compared only through the residual detrended process. First, this component is modelled by directly specifying certain spatio-temporal correlation functions with increasing complexity, from the i.i.d. over time case to the nonseparable form (Model A1, A2, A3-1 and A3-2). Then, in Model B the detrended process is defined as the sum of a constant in space process, with an AR(1) temporal dynamics, and a purely spatial process. Model C, on the other hand, features a spatial process which evolves in time according to an AR(1) equation with innovations i.i.d. over time. 

Our case study, concerning the prediction of PM$_{10}$ in Piemonte, allows us to identify the best combination of hierarchical structure and spatio-temporal covariance function based on the evaluation of the proposed criteria. The application highlights that complex spatio-temporal covariances (in Model A3-1 and A3-2) require computational costs which are too high with respect to the observed prediction capability. The separable Model A2 is discarded because of its poor performance in the predicting procedure (one star). Model B, on the other hand, is comparable to Model A1 with respect to the prediction capability (two stars) but it is slower from a computational point of view. Model C is the only three-star predictor: this suggests that, in our case study, a model with a complex hierarchical structure is globally preferable to one with a complex spatio-temporal covariance function. 
Thus, the last comparison is between Model A1 and C: since the latter requires small additional computational cost in order to be awarded an additional star, our final suggestion is to choose according to the available computational resources. 

Obviously, the conclusions and the final suggestion in our case study do not constitute the general answer to the question about the best combination of hierarchical structure and spatio-temporal covariance function. For example, Model A3-1 and A3-2 could have a better prediction performance in a case study with a larger spatial domain and a longer time period, as well as a thicker monitoring network. Moreover, their computational costs could be reduced by simplifying the structure of the dense covariance matrices, using for example covariance tapering techniques \citep{Tapering}. Nonetheless, the proposed set of criteria summarized in a simple table - jointly with the model discussion - makes the strategy applied in our PM$_{10}$ application portable to other environmental studies, where decision makers choose which model to implement.

The choice of the model is also related to the size of the dataset used to estimate the parameters as well as the grid resolution for spatial mapping, since we adopt computing intensive statistical methods. The increasing availability of large spatio-temporal datasets in many environmental fields gives rise to the so-called ``big n'' problem and to the infeasibility of matrix operations whose complexity increases in cubic order. To tackle this issue, we need to adopt efficient computational strategies. A solution may be given by the implementation of parallel linear algebra algorithms based on graphical processors (GPU), which is part of our ongoing research.

\section*{Acknowledgements}
The authors would like to thank Veronica J. Berrocal, Alan E. Gelfand, Giovanna Jona Lasinio for the useful discussions and comments. 

\bibliographystyle{natbib}
\bibliography{biblio}

\begin{thebibliography}{}

\bibitem[Al-Awadhi and Al-Awadhi(2006)Al-Awadhi and Al-Awadhi]{kuwait}
Al-Awadhi, F. and Al-Awadhi, S. (2006).
\newblock Spatial-temporal model for ambient air pollutants in the state of
  {K}uwait.
\newblock {\em Environmetrics\/}, {\bf 17}, 739--752.

\bibitem[Bande {\em et~al.}(2007)Bande, Clemente, De~Maria, Muraro, Picollo,
  Arduino, Calori, Finardi, Radice, Silibello, and Brusasca]{Bande}
Bande, S., Clemente, M., De~Maria, R., Muraro, M., Picollo, M., Arduino, G.,
  Calori, G., Finardi, S., Radice, P., Silibello, C., and Brusasca, G. (2007).
\newblock The modelling system supporting {P}iemonte region yearly air quality
  assessment.
\newblock {\em Proceedings of 6th International Conference on Urban Air
  Quality, Limassol, Cyprus, 27-29 March 2007\/}.

\bibitem[Banerjee {\em et~al.}(2004)Banerjee, Carlin, and
  Gelfand]{hierarchicalmodeling}
Banerjee, S., Carlin, B., and Gelfand, A. (2004).
\newblock {\em Hierarchical Modeling and Analysis for Spatial Data\/}.
\newblock Monographs on Statistics and Applied Probability. Chapman and Hall,
  New York.

\bibitem[Clark(2005)Clark]{why}
Clark, J. (2005).
\newblock Why environmental scientists are becoming {B}ayesians.
\newblock {\em Ecology Letters\/}, {\bf 8}, 2--14.

\bibitem[Cocchi {\em et~al.}(2007)Cocchi, Greco, and Trivisano]{cocchi}
Cocchi, D., Greco, F., and Trivisano, C. (2007).
\newblock Hierarchical space-time modelling of {PM}$_{10}$ pollution.
\newblock {\em Atmospheric environment\/}, {\bf 41}, 532--542.

\bibitem[Cressie(1993)Cressie]{cressie93}
Cressie, N. (1993).
\newblock {\em Statistics for Spatial Data\/}.
\newblock Wiley, New York.

\bibitem[Cressie and Huang(1999)Cressie and Huang]{huangcressie}
Cressie, N. and Huang, H. (1999).
\newblock Classes of nonseparable, spatio-temporal stationary covariance
  functions.
\newblock {\em Journal of the American Statistical Association\/}, {\bf 94},
  1330--1340.

\bibitem[Fass{\`o} and Cameletti(2010)Fass{\`o} and Cameletti]{simulation08}
Fass{\`o}, A. and Cameletti, M. (2010).
\newblock A unified statistical approach for simulation, modelling, analysis
  and mapping of environmental data.
\newblock {\em Simulation\/}, {\bf 86}(3), 139--154.

\bibitem[Fass\`{o} {\em et~al.}(2007)Fass\`{o}, Cameletti, and Nicolis]{GDC}
Fass\`{o}, A., Cameletti, M., and Nicolis, O. (2007).
\newblock Air quality monitoring using heterogeneous networks.
\newblock {\em Environmetrics\/}, {\bf 18}(3), 245--264.

\bibitem[Finardi {\em et~al.}(2008)Finardi, De~Maria, D'Allura, Cascone,
  Calori, and Lollobrigida]{deterministicmodels}
Finardi, S., De~Maria, R., D'Allura, A., Cascone, C., Calori, G., and
  Lollobrigida, F. (2008).
\newblock A deterministic air quality forecasting system for {T}orino urban
  area, italy.
\newblock {\em Environmental Modelling and Software\/}, {\bf 23}(3), 344--355.

\bibitem[Fuentes(2006)Fuentes]{fuentesseparability}
Fuentes, M. (2006).
\newblock Testing for separability of spatial-temporal covariance functions.
\newblock {\em Journal of Statistical Planning and Inference\/}, {\bf 136},
  447--466.

\bibitem[Furrer {\em et~al.}(2006)Furrer, Genton, and Nychka]{Tapering}
Furrer, R., Genton, M., and Nychka, D. (2006).
\newblock Covariance tapering for interpolation of large spatial datasets.
\newblock {\em Journal of Computational and Graphical Statistics\/}, {\bf
  15}(3), 502--523.

\bibitem[Gneiting(2002)Gneiting]{Gneiting}
Gneiting, T. (2002).
\newblock Nonseparable, stationary covariance functions for space-time data.
\newblock {\em Journal of the American Statistical Association\/}, {\bf 97},
  590--600.

\bibitem[Harrison {\em et~al.}(2008)Harrison, Stedman, and
  Derwent]{whynotfalling}
Harrison, R., Stedman, J., and Derwent, D. (2008).
\newblock New directions: {W}hy are {PM}$_{10}$ concentrations in {E}urope not
  falling?
\newblock {\em Atmospheric Environment\/}, {\bf 42}, 603--606.

\bibitem[Huang {\em et~al.}(2007)Huang, Martinez, Mateu, and
  Montes]{huangcomparison}
Huang, H., Martinez, F., Mateu, J., and Montes, F. (2007).
\newblock Model comparison and selection for stationary space-time models.
\newblock {\em Computational Statistics and Data Analysis\/}, {\bf 51},
  4577--4596.

\bibitem[Kolovos {\em et~al.}(2006)Kolovos, Yu, and Christakos]{SEKSGUI}
Kolovos, A., Yu, H., and Christakos, G. (2006).
\newblock {\em SEKS-GUI v.0.6 User Manual\/}.
\newblock Department of Geography, San Diego State University: San Diego, CA.

\bibitem[Pang {\em et~al.}(2010)Pang, Christakos, and Wang]{pangcomparison}
Pang, W., Christakos, G., and Wang, J. (2010).
\newblock Comparative spatiotemporal analysis of fine particulate matter
  pollution.
\newblock {\em Environmetrics\/}, {\bf 21}, 305--317.

\bibitem[Poli and Cirillo(1993)Poli and Cirillo]{Poli}
Poli, A. and Cirillo, M. (1993).
\newblock On the use of the normalized mean square error in evaluating
  dispersion model performance.
\newblock {\em Atmospheric Environment\/}, {\bf 27}, 2427--2434.

\bibitem[Pollice and Jona~Lasinio(2010a)Pollice and
  Jona~Lasinio]{GRASPA32_publ}
Pollice, A. and Jona~Lasinio, G. (2010a).
\newblock A multivariate approach to the analysis of air quality in a high
  environmental risk area.
\newblock {\em Environmetrics\/}.
\newblock DOI: 10.1002/env.1059.

\bibitem[Pollice and Jona~Lasinio(2010b)Pollice and
  Jona~Lasinio]{GRASPA31_publ}
Pollice, A. and Jona~Lasinio, G. (2010b).
\newblock Spatiotemporal analysis of the {PM}$_{10}$ concentration over the
  {T}aranto area.
\newblock {\em Environmental Monitoring and Assessment\/}, {\bf 162}, 177--190.

\bibitem[Sahu and Mardia(2005)Sahu and Mardia]{bayesiankkf}
Sahu, S. and Mardia, K. (2005).
\newblock A {B}ayesian {K}riged-{K}alman model for short-term forecasting of
  air pollution level.
\newblock {\em Journal of the Royal Statistical Society, Series C\/}, {\bf 54},
  223--244.

\bibitem[Sahu and Nicolis(2007)Sahu and Nicolis]{sahunicolis}
Sahu, S. and Nicolis, O. (2007).
\newblock An evaluation of {E}uropean air pollution regulations for particulate
  matter monitored from a heterogeneous network.
\newblock {\em Environmetrics\/}, {\bf 20}, 943--961.

\bibitem[Sahu {\em et~al.}(2006)Sahu, Gelfand, and Holland]{STfinePM}
Sahu, S., Gelfand, A., and Holland, D. (2006).
\newblock Spatio-temporal modeling of fine particulate matter.
\newblock {\em Journal of Agricultural, Biological and Environmental
  Statistics\/}, {\bf 11}, 61--86.

\bibitem[Shaddick and Wakefield(2002)Shaddick and Wakefield]{shaddick}
Shaddick, G. and Wakefield, J. (2002).
\newblock Modelling daily multivariate pollutant data at multiple sites.
\newblock {\em Journal of Applied Statistics\/}, {\bf 3}, 351--372.

\bibitem[Smith and Kolenikov(2003)Smith and Kolenikov]{pm2.5smith}
Smith, R. and Kolenikov, S. (2003).
\newblock Spatiotemporal modeling of {PM}$_{2.5}$ data with missing values.
\newblock {\em Journal of Geophysical Resarch\/}, {\bf 108}(D24), 11--1,11--11.

\bibitem[Tanner(1996)Tanner]{composition}
Tanner, M.~A. (1996).
\newblock {\em Tools for Statistical Inference: Methods for the Exploration of
  Posterior Distributions and Likelihood Functions\/}.
\newblock Springer, New York.

\bibitem[West and Harrison(1997)West and Harrison]{FFBS}
West, M. and Harrison, J. (1997).
\newblock {\em Bayesian Forecasting and Dynamic Models\/}.
\newblock Springer.

\bibitem[Wikle(2003)Wikle]{hierarchical}
Wikle, C.~K. (2003).
\newblock Hierarchical models in environmental science.
\newblock {\em International Statistical Review\/}, {\bf 71}, 181--199.

\bibitem[Wikle {\em et~al.}(1998)Wikle, Berliner, and
  Cressie]{hierarchical_spacetime}
Wikle, C.~K., Berliner, L., and Cressie, N. (1998).
\newblock Hierarchical bayesian space-time models.
\newblock {\em Journal of Environmental and Ecological Statistics\/}, {\bf 5},
  117--154.

\bibitem[Yu {\em et~al.}(2006)Yu, Eder, Dennis, Chu, and Schwartz]{Yu}
Yu, S., Eder, B., Dennis, R., Chu, S., and Schwartz, S. (2006).
\newblock New unbiased symmetric metrics for evaluation of air quality models.
\newblock {\em Atmospheric Science Letters\/}, {\bf 7}, 26--34.

\bibitem[Zidek {\em et~al.}(2002)Zidek, Sun, Le, and Ozkaynak]{zidek}
Zidek, J., Sun, L., Le, N., and Ozkaynak, H. (2002).
\newblock Contending with space-time interaction in the spatial prediction of
  pollution: Vancouver's hourly ambient {PM}$_{10}$ field.
\newblock {\em Environmetrics\/}, {\bf 13}, 595--613.

\end{thebibliography}

\appendix
\newpage

\section{Details on covariance separability}\label{app0}
For Model B it holds that
\begin{eqnarray*}
Cov\left(u\left(s_i,t\right),u\left(s_j,t^\prime\right)\right)&=&E\left(u\left(s_i,t\right)u\left(s_j,t^\prime\right)\right)-E\left(u\left(s_i,t\right)\right)E\left(u\left(s_j,t^\prime\right)\right)\\
&=&E\left(y(t)y(t^\prime)\right)-E\left(y(t)\right)E\left(y(t^\prime)\right)+E\left(\omega(s_i)\omega(s_j)\right)\\
&=&Cov\left(y\left(t\right),y\left(t^\prime\right)\right)+Cov\left(\omega(s_i),\omega(s_j)\right)\\
&=&\rho^l \frac{\sigma^2_\eta}{1-\rho^2}+\sigma^2_\omega C_{\theta}(h)
\end{eqnarray*}
%
where $h=\left\|s_i-s_j\right\|$ and $l=\left|t-t^\prime\right|$. This means that the spatio-temporal covariance function is given by the sum of a purely temporal and a purely spatial covariance function. In particular, the temporal term is defined by the AR(1) structure of $y$ (see \eqref{modelBeq3}) with $|\rho|<1$. 

Turning to Model C we have that $Cov\left(u\left(s_i,t\right),u\left(s_j,t^\prime\right)\right)$ is equal to $Cov\left(y\left(s_i,t\right),y\left(s_j,t^\prime\right)\right)$. Moreover, recalling that any AR(1) process can be rewritten using an infinite-order moving average representation, it follows that 
 \begin{eqnarray*}
Cov\left(y\left(s_i,t\right),y\left(s_j,t^\prime\right)\right)&=&Cov\left(\sum_{k=0}^\infty \rho^k \omega(s_i,t-k),\sum_{k^\prime=0}^\infty \rho^{k^\prime} \omega(s_j,t^\prime-k^\prime)\right)\\
&=&\sum_{k,k^\prime=0}^\infty \left(\rho^{k+k^\prime} Cov\left(\omega(s_i,t-k),\omega(s_j,t^\prime-k^\prime)\right)\right)\\
&=&  \frac{\rho^l}{1-\rho^2}\sigma^2_\omega C_{\theta}(h).
 \end{eqnarray*}
Hence the spatio-temporal covariance function can be rewritten in a separable multiplicative form where $\frac{\rho^l}{1-\rho^2}$ is a valid temporal correlation function if $|\rho|<1$.

\section{Derivation of posterior distributions}\label{app1}

Let us denote with $\Psi$ the parameter vector to be estimated. Generally speaking the joint posterior distribution is given by
\begin{equation}\label{eq:joint1}
f\left(\Psi, U\mid Z\right) \propto f\left(Z \mid U,\Psi\right)f(U\mid\Psi)f(\Psi)
\end{equation}
where the notation $f(.)$ is used for the probability density function and $Z$ and $U$ denote, respectively, the collections of data $z(s_i,t)$ and of the latent processes $u\left(s_{i},t\right)$. Note that independent prior distributions are chosen for the parameters, so that $f(\Psi)=\prod_{i=1}^{dim(\Psi)}f(\Psi_i)$.
Conditionally on $U$ the observations $z\left(s_{i},t\right)$ are serially independent and Eq.\eqref{eq:joint1} can be rewritten as follows:
\begin{equation}\label{eq:joint2}
f\left(\Psi,U\mid Z\right) \propto \prod_{t=1}^{T}f\left(Z_t\mid U,\Psi\right)f(U\mid\Psi)f(\Psi)
\end{equation}
where $Z_t=\left(z(s_1,t),\ldots,z(s_d,t)\right)^\prime$ and $f\left(Z_t\mid U,\Psi\right)$ can be derived through the measurement equation \eqref{meas.eq}. The joint posterior distribution \eqref{eq:joint2} is completely specified when $f(U\mid\Psi)$ is defined. Since this conditional distribution is specific for each model introduced in Section 2, we describe each case separately in the following subsections specifying in the subscript of $\Psi$ the model to which it refers.

Note that, as described in \cite{hierarchicalmodeling}, in the case of hierarchical models it is preferable, where possible, to marginalize the posterior distribution over $U$, thus obtaining the following posterior distribution
\begin{equation}\label{eq:joint3}
f\left(\Psi\mid Z\right) \propto \prod_{t=1}^{T}f\left(Z_t\mid \Psi\right)f(\Psi).
\end{equation}
In the sequel, this marginal approach is used for all models except Model A2, where the marginalization prevents the use of the properties of the Kronecker product described in Section \ref{sec:A2}. 

In order to implement the Gibbs sampling algorithm we derive the full conditional distributions and, when these are not available in an exact closed-form, we introduce a Metropolis-Hastings (MH) sampling step (which is the so-called Metropolis-within-Gibbs algorithm).

\subsection{Model A1}\label{sec:A1}
As specified in Section 2.1 the parameter vector for Model A1 is given by $\Psi_{A_1}=\left(\beta, \sigma^2_\varepsilon, \theta,\sigma^2_\omega\right)$. 
Moreover, the following Gaussian $d$-dimensional conditional distribution holds
\begin{equation*}
f\left(Z_t\mid \Psi_{A_1}\right) \sim N_d\left(X_t\beta,\Sigma_{\omega+\varepsilon}\right)
\end{equation*}
where $X_t=\left(X(s_1,t)^\prime,\ldots,X(s_d,t)^\prime\right)^\prime$ is the $(d\times k)$ covariate matrix and the variance-covariance matrix is defined as 
\begin{equation}
\Sigma_{\omega+\varepsilon}=\sigma^{2}_{\omega}C_\theta\left(\|s_i-s_j\|\right)_{i,j=1,\ldots,d}+\sigma^{2}_{\varepsilon}I_d, \label{eq:Sigmaomegavarepsilon}
\end{equation}
with $C_\theta\left(\|s_i-s_j\|\right)_{i,j=1,\ldots,d}=C_\theta(\mathbf{h})$ with elements evaluated through the spatial correlation function shown in (2.5). Then, it follows that the posterior distribution is given by
\begin{eqnarray*}
f\left(\Psi_{A1}\mid Z\right)&\propto&\prod_{i=1}^{dim(\Psi_{A1})}f({\Psi_{A1}}_i)\\
&\times&\left|\Sigma_{\omega+\varepsilon}\right|^{-\frac{T}{2}}\exp\left[-\frac{1}{2}\sum^{T}_{t=1}\left(Z_t-X_t\beta\right)^{\prime}\Sigma_{\omega+\varepsilon}^{-1}\left(Z_t-X_t\beta\right)\right].\\
\end{eqnarray*}

Taking a Normal prior distribution for $\beta$, that is $\beta \sim N_k(0,\Sigma_0)$, straightforward calculation yields a Gaussian full conditional distribution for $\beta$ with mean $AB^\prime$ and covariance A where $A =\left(\Sigma_0^{-1}+\sum^{T}_{t=1}X_t^{\prime}\Sigma_{\omega+\varepsilon}^{-1}X_t\right)^{-1}$ and
$B = \sum^{T}_{t=1}\left(Z_t^{\prime}\Sigma_{\omega+\varepsilon}^{-1}X_t\right)$. In order to estimate the remaining parameters, the MH algorithm is used.

\subsection{Model A2}\label{sec:A2}
Let us denote with $U_t=\left(U(s_1,t),\ldots,U(s_d,t)\right)^\prime$ the latent process at time $t$. Moreover, let $U=\{U_1,\ldots,U_T\}$ be the $(dT\times 1)$ random effects vector blocked by sites\footnote{Here, and in the sequel, braces are used for column stacking of the vectors involved.} and $Z=\{Z_1,\ldots,Z_T\}$ the $(dT\times 1)$ data vector. The distribution of $U$ is Gaussian with mean vector given by $X\beta$ where $X=\{X_1,\ldots,X_T\}$ is the $(dT\times k)$ array of covariates and $\beta$ is the corresponding coefficient vector. The variance-covariance matrix of $U$ is defined as
\[
\Sigma_U=\sigma^{2}_{\omega}C_{\theta}\left(\|s_i-s_j\|,|t-t^\prime|\right)_{\substack{i,j=1,\ldots,d \\ t,t^\prime=1,\ldots,T}}=\sigma^{2}_{\omega}C_{\theta}\left(\mathbf{h},\mathbf{l}\right)
\]
where $C_{\theta}\left(\|s_i-s_j\|,|t-t^\prime|\right)_{\substack{i,j=1,\ldots,d \\ t,t^\prime=1,\ldots,T}}=C_{\theta}\left(\mathbf{h},\mathbf{l}\right)$ with elements evaluated through the separable spatio-temporal correlation function given in Eq.\eqref{eq:sepCov}.


Thanks to the separability property, the correlation matrix $C_{\theta}\left(\mathbf{h},\mathbf{l}\right)$ can be written as $C_{\theta}\left(\mathbf{h},\mathbf{l}\right)=C_{\theta_1}\left(\mathbf{l}\right)\otimes C_{\theta_2}\left(\mathbf{h}\right)$, where $C_{\theta_1}\left(\mathbf{l}\right)$ and $C_{\theta_2}\left(\mathbf{h}\right)$ denote respectively the temporal and spatial correlation matrix. Thus, the parameter vector for Model A2 is $\Psi_{A2}=\left(\beta,\sigma^{2}_{\varepsilon},\sigma^{2}_{\omega},\theta_1,\theta_2\right)$ and the joint posterior distribution is given by the following equation:
\begin{eqnarray*}\label{eq:jointA2}
f\left(\Psi_{A2},U\mid Z\right) &\propto&\left(\sigma^{2}_{\varepsilon}\right)^{-\frac{Td}{2}}\exp\left[-\frac{1}{2\sigma^{2}_{\varepsilon}}\left(Z-U\right)^\prime\left(Z-U\right)\right]\nonumber\\
&\times&\left(\sigma^{2}_{\omega}\right)^{-\frac{Td}{2}}
\left|C_{\theta}(\mathbf{h},\mathbf{l})\right|^{-\frac{1}{2}}\exp\left[-\frac{1}{2\sigma^{2}_{\omega}}(U-X\beta)^{\prime}C_{\theta}(\mathbf{h},\mathbf{l})^{-1}(U-X\beta)\right]\\
&\times& \prod_{i=1}^{dim(\Psi_{A2})}f({\Psi_{A2}}_i)\nonumber
\end{eqnarray*}
where, thanks to the Kronecker product properties, $\left|C_{\theta}(\mathbf{h},\mathbf{l})\right|=\left|C_{\theta_1}(\mathbf{l})\right|^{d}\left|C_{\theta_2}(\mathbf{h})\right|^{T}$ and $C_{\theta}(\mathbf{h},\mathbf{l})^{-1}=C_{\theta_1}(\mathbf{l})^{-1}\otimes C_{\theta_2}(\mathbf{h})^{-1}$.

Taking a Normal prior distribution for $\beta$, that is $\beta \sim N_k(0,\Sigma_0)$, from Eq.\eqref{eq:jointA2} it follows that the full conditional distribution for $\beta$ is Gaussian with mean $AB^\prime$ and variance $A$ where $A = \left(\Sigma_0^{-1}  + \frac{1}{\sigma^2_\omega}X^\prime C_{\theta}(\mathbf{h},\mathbf{l})^{-1} X\right)^{-1}$ and $B =\frac{1}{\sigma^2_\omega}U^\prime C_{\theta}(\mathbf{h},\mathbf{l})^{-1} X$. Moreover, for the variance parameter we have the following conditional distributions:
\[
\sigma^2_\omega \sim IG\left(a+\frac{Td}{2},b+\frac{1}{2}(U-X\beta)^\prime C_{\theta}(\mathbf{h},\mathbf{l})^{-1}(U-X\beta)\right),
\]
\[
\sigma^2_\varepsilon \sim IG\left(a+\frac{Td}{2},b+\frac{1}{2}\left(Z-U\right)^\prime\left(Z -U\right)\right)
\]
where $a$ and $b$ denote the hyperparameters of the corresponding prior Inverse-Gamma distributions $IG(a,b)$. For the estimation of $\theta_1$ and $\theta_2$ we adopt the MH algorithm. Finally, let us recall that the spatio-temporal process $U$ is sampled using the \emph{en bloc} procedure described in \citet{STfinePM}.

\subsection{Model A3}\label{sec:A3}
Using the notation introduced in the previous section for Model A2, the posterior distribution for Model A3 is
\begin{equation}\label{eq:jointA3}
f\left(\Psi_{A3}\mid Z\right) \propto |\Sigma_{\omega+\varepsilon}|^{-\frac{1}{2}}\exp\left[-\frac{1}{2}\left(Z-X\beta\right)^\prime\Sigma_{\omega+\varepsilon}^{-1}\left(Z-X\beta\right)\right] \times \prod_{i=1}^{dim(\Psi_{A3})}f({\Psi_{A3}}_i)
\end{equation}
where $\Psi_{A3-1}=\left(\beta,\sigma^{2}_{\varepsilon},\sigma^{2}_{\omega},\theta_{A3-1}\right)$ for Model A3-1 and $\Psi_{A3-2}=\left(\beta,\sigma^{2}_{\varepsilon},\sigma^{2}_{\omega},\theta_{A3-2}\right)$ for Model A3-2. The variance-covariance matrix in \eqref{eq:jointA3} is given by
\[
\Sigma_{\omega+\varepsilon}=\sigma^{2}_{\omega} C_\theta\left(\mathbf{h},\mathbf{l}\right)+\sigma^{2}_{\varepsilon}I_{dT}
\]
where the elements of $C_\theta\left(\mathbf{h},\mathbf{l}\right)$ are evaluated through the nonseparable correlation functions $C_\theta(h,l)$ given in Eq. (2.7) of \eqref{covGNE}.

From Eq.\eqref{eq:jointA3} we can easily obtain that the full conditional distribution for $\beta$ is Gaussian with mean $AB^\prime$ and variance $A$ where $A =\left( \Sigma_0^{-1}  + X^\prime \Sigma_{\omega+\varepsilon}^{-1} X\right)^{-1}$ and $B =Z^\prime \Sigma_{\omega+\varepsilon}^{-1} X$,
assuming a Normal prior distribution for $\beta$, that is $\beta \sim N_k(0,\Sigma_0)$. All the remaining parameters are estimated through the Metropolis-Hastings algorithm.
\subsection{Model B}\label{sec:B}
Given the hierarchical structure of Model B (see Equations \eqref{modelBeq2} and \eqref{modelBeq3}), Eq.\eqref{eq:joint3} can be rewritten as 
\begin{equation}\label{eq:jointB}
f\left(\Psi_{B}\mid Z \right)=\prod_{t=1}^T f\left(Z_t \mid Y, \Psi_B\right) f\left(Y\mid \Psi_B\right) f\left(\Psi_B\right)
\end{equation}
where $\Psi_{B}=\left(\beta,\sigma^{2}_{\varepsilon},\sigma^{2}_{\omega},\sigma^{2}_{\eta},\theta,\rho\right)$ and $Y=\{Y_{1},\ldots,Y_{T}\}$, with $Y_t$ denoting the constant in space unidimensional latent process at time $t$, with $ t=1, \ldots, T$. The Markovian structure defined in Eq.\eqref{modelBeq3} leads us to rewrite the term $f\left(Y\mid\Psi_{B}\right)$ in \eqref{eq:jointB} as follows
\begin{equation}\label{eq:jointB2}
f\left(Y\mid\Psi_{B}\right)=f\left(Y_{0}\mid\Psi_{B}\right)\prod_{t=1}^{T}f\left(Y_{t}\mid Y_{t-1},\Psi_{B}\right).
\end{equation}
Moreover, the conditional probability distributions that occur in the previous equations are the following:
\begin{eqnarray*}
f\left(Z_{t}\mid Y,\Psi_{B}\right)&\sim&N_{d}\left(X_{t}\beta+K Y_{t},\Sigma_{\omega+\varepsilon}\right),\\
f\left(Y_{t}\mid Y_{t-1},\Psi_{B}\right)&\sim&N_1\left(\rho Y_{t-1},\sigma^{2}_{\eta}\right),\\
f\left(Y_{0}\mid \Psi_{B}\right)&\sim&N_1\left(0,\sigma^{2}_{B}\right),
\end{eqnarray*}
where $\Sigma_{\omega+\varepsilon}$ is defined by Eq.\eqref{eq:Sigmaomegavarepsilon}, as for Model A1, and $K$ is a $(d\times 1)$-dimensional ones vector. 
Taking the logarithm of Eq. \eqref{eq:jointB}, the posterior distribution for Model B is given by
\begin{eqnarray*}\label{eq:jointB3}
f\left(\Psi_{B}\mid Z\right) &\propto&
|\Sigma_{\omega+\varepsilon}|^{-\frac{1}{2}}\exp\left[-\frac{1}{2}\sum_{t=1}^T\left(Z_t-X_t\beta-KY_t\right)^\prime \Sigma_{\omega+\varepsilon}^{-1}\left(Z_t-X_t\beta-KY_t\right)\right]\\
&\times&\left(\sigma^{2}_{B}\right)^{-\frac{1}{2}}\exp\left(-\frac{1}{2}Y_0^{2}\right)\\
&\times&\left(\sigma^{2}_{\eta}\right)^{-\frac{T}{2}}\exp\left[-\frac{1}{2\sigma^{2}_{\eta}}\sum^{T}_{t=1}\left(Y_t-\rho Y_{t-1}\right)^{2}\right]\prod_{i=1}^{dim(\Psi_B)}f({\Psi_{B}}_i).
\end{eqnarray*}

Straightforward calculation yields a Gaussian full conditional distribution for $\beta$ with mean $AB^\prime$ and covariance A where $A =\left(\Sigma_0^{-1}+\sum^{T}_{t=1}X_t^{\prime}\Sigma_{\omega+\varepsilon}^{-1}X_t\right)^{-1}$ and $B = \sum^{T}_{t=1}\left(Z_t^{\prime}\Sigma_{\omega+\varepsilon}^{-1}X_t-(KY_t)^{\prime}\Sigma_{\omega+\varepsilon}^{-1}X_t\right)$, 
assuming a Normal prior distribution for $\beta$, that is $\beta \sim N_k(0,\Sigma_0)$. For the estimation of $\sigma^2_\eta$ we consider an Inverse Gamma prior $IG(a,b)$ and sample from the following full conditional distribution
\[
\sigma^{2}_{\eta} \sim IG \left(\frac{T}{2}+a,b+\frac{1}{2}\sum^{T}_{t=1}\left(Y_t- \rho Y_{t-1}\right)^{2}\right).
\]

For the other parameters, $\sigma^2_\varepsilon,\sigma^2_\omega,\theta$ and $\rho$, we adopt the MH algorithm, while the latent process $Y$ is sampled using the Forward Filtering Backward Sampling (FFBS) as described in \cite{FFBS}.
\subsection{Model C}\label{sec:C}
With regard to Model C, Eq.\eqref{eq:jointB} and Eq.\eqref{eq:jointB2} still hold substituting $\Psi_B$ with $\Psi_C$ and using the following conditional distributions:
\begin{eqnarray*}
f\left(Z_{t}\mid Y,\Psi_{C}\right)&\sim&N_{d}\left(X_{t}\beta+Y_{t},\Sigma_{\varepsilon}\right),\\
f\left(Y_{t}\mid Y_{t-1},\Psi_{C}\right)&\sim&N_d\left(P Y_{t-1},\Sigma_{\omega}\right),\\
f\left(Y_{0}\mid \Psi_{C}\right)&\sim&N_d\left(0,\Sigma_{0}\right),
\end{eqnarray*}
where $Y_t=\left(Y(s_1,t),\ldots,Y(s_d,t)\right)^\prime$ and $P=\rho I_d$ is the $AR(1)$ transition matrix, $\Sigma_{\varepsilon}=\sigma^2_\varepsilon I_d$ and $\Sigma_\omega=\sigma^2_\omega C_\theta\left(\|s_i-s_j\|\right)_{i,j=1,\ldots,d}=\sigma^2_\omega C_\theta(\mathbf{h})$, with elements evaluated through the spatial correlation function appearing in \eqref{iidCov}. It follows that the posterior distribution is
\begin{eqnarray*}\label{eq:jointC}
f\left(\Psi_{C}\mid Z\right) &\propto& \left(\sigma^2_\varepsilon\right)^{-\frac{dT}{2}}\exp\left[-\frac{1}{2\sigma^2_\varepsilon}\sum_{t=1}^T\left(Z_t-X_t\beta-Y_t\right)^\prime\left(Z_t-X_t\beta-Y_t\right) \right]\\
&\times&\left|\Sigma_0\right|^{-\frac{1}{2}}\exp\left[-\frac{1}{2}Y_0^{\prime}\Sigma^{-1}_{0}Y_0\right]\\
&\times&\left(\sigma^{2}_{\omega}\right)^{-\frac{dT}{2}}\left|C_{\theta}(\mathbf{h})\right|^{-\frac{T}{2}}\exp\left[-\frac{1}{2\sigma^{2}_{\omega}}\sum^{T}_{t=1}\left(Y_t- P Y_{t-1}\right)^{\prime}C_{\theta}(\mathbf{h})^{-1}\left(Y_t-P Y_{t-1}\right)\right]\\
&\times&\prod_{i=1}^{dim(\Psi_C)}f({\Psi_{C}}_i).
\end{eqnarray*}

Straightforward calculation yields a Gaussian full conditional distribution for $\beta$ with mean $AB^\prime$ and covariance A where
$A =\left(\Sigma_0^{-1}+\frac{1}{\sigma^{2}_{\varepsilon}}\sum^{T}_{t=1}X_t^{\prime}X_t\right)^{-1}$ and
$B = \frac{1}{\sigma^{2}_{\varepsilon}}\sum^{T}_{t=1}\left(Z_t^{\prime}X_t-Y_t^{\prime}X_t\right)$, assuming a Normal 
prior distribution for $\beta$, that is $\beta \sim N_k(0,\Sigma_0)$. Moreover, for the variance parameters we have the following conditional posterior distributions:
\begin{eqnarray*}
\sigma^2_\omega &\sim& IG\left(\frac{dT}{2}+a,b+\frac{1}{2}\sum^{T}_{t=1}\left(Y_t-P Y_{t-1}\right)^{\prime}C_{\theta}(\mathbf{h})^{-1}\left(Y_t-P Y_{t-1}\right)\right),\\
\sigma^2_\varepsilon &\sim& IG\left(\frac{dT}{2}+a,b+\frac{1}{2}\sum^{T}_{t=1}\left(Z_t-X_t\beta-Y_t\right)^{\prime}\left(Z_t-X_t\beta-Y_t\right)\right),
\end{eqnarray*}
where $a$ and $b$ denote the hyperparameters of the corresponding Inverse-Gamma priors $IG(a,b)$. The parameters $\theta$ and $\rho$ are estimated using the MH algorithm while latent process $Y$ is sampled using the FFBS algorithm.
\section{Spatial prediction}\label{app2}
Spatial prediction at a new location $s_0$ and time $t_0$ (with $1\leq t_0 \leq T$) is based on the posterior predictive distribution of $z(s_0,t_0)$ which is given by
\begin{equation}\label{eq:posteriorpredictive}
f\left(z(s_0,t_0)\mid Z\right) =\int f\left(z(s_0,t_0),u(s_0,t_0),\Psi \mid Z\right) d u(s_0,t_0)d\Psi.
\end{equation}

Note that for all the models described in Section \ref{sec:models} except Model A2, we can marginalize over $U$ and \eqref{eq:posteriorpredictive} reduces to the following equation
\begin{equation}\label{eq:posteriorpredictive2}
f\left(z(s_0,t_0)\mid Z\right) = \int f\left(z(s_0,t_0),\Psi \mid Z\right) d\Psi =\int f\left(z(s_0,t_0)\mid Z, \Psi\right)f\left(\Psi \mid Z\right) d\Psi.
\end{equation}

In practice, in order to obtain a prediction using MCMC methods, the posterior predictive distribution \eqref{eq:posteriorpredictive2} is sampled by composition \citep{composition}. This means that we first draw from the posterior $f\left(\Psi \mid Z\right) $ which then allows us to draw from $f\left(z(s_0,t_0)\mid Z, \Psi\right)$. Therefore, to define $f\left(z(s_0,t_0)\mid Z, \Psi\right)$ we refer to the standard multivariate Normal theory. In fact, from the joint distribution of the data $Z$ and $z(s_0,t_0)$, that is
\[
\left(
\begin{array}
[c]{c}%
Z\\
z\left(  s_{0},t_0\right)
\end{array}
\bigg| \bigg. \Psi\right)  \sim N_{dT+1}\left[  \left(
\begin{array}
[c]{c}%
\mu_{1}\\
\mu_{2}%
\end{array}
\right)  ,\left(
\begin{array}
[c]{cc}%
\Sigma_{11} & \Sigma_{12}\\
\Sigma_{12}^{\prime} & \Sigma_{22}
\end{array}
\right)  \right],
\]
it derives that 
\begin{equation}\label{eq:cond posterior}
z\left(s_0,t_0\right) \mid Z, \Psi \sim N_1 \left(\mu_2+\Sigma_{12}^\prime\Sigma_{11}^{-1}\left(Z-\mu_1\right), \Sigma_{22}-\Sigma_{12}^\prime\Sigma_{11}^{-1}\Sigma_{12}\right).
\end{equation}
Note that  $\Sigma_{11}$, the variance-covariance matrix of the data $Z$, has dimension $(dT\times dT)$ while $\Sigma_{22}$ is a scalar. Moreover, the covariance vector $\Sigma_{12}$ has dimension $(dT\times 1)$ and its generic element is $Cov\left(
z\left( s_{i},t\right)  ,z\left(  s_{0},t_0\right)  \right)$ $(i=1,\ldots,d;t=1,\ldots,T)$. In the following subsections we report schematically for each model the elements $\mu_1$, $\mu_2$, $\Sigma_{11}$, $\Sigma_{22}$ and $\Sigma_{12}$ of \eqref{eq:cond posterior} referring to the notation introduced in Section \ref{app1}. Moreover, let $\mathbf{\tilde h}$ denote the vector of distances $\|s_i-s_0\|$ with $i=1,\ldots,d$, and $\mathbf{\tilde l}$ the vector of temporal lags $\|t-t_0\|$ with $t=1,\ldots,T$. Thus, for example, the term $C_\theta\left(\mathbf{\tilde h}\right)$ is the correlation vector whose generic element is $C_\theta\left(\|s_i-s_0\|\right)$, with $i=1,\ldots,d$.

\subsection{Model A1}\label{sec:A1SP}
 For Model A1 it holds that $\mu_1=X\beta$, where $X=\{X_1,\ldots,X_T\}$ is the $(dT\times k)$ array of covariates, and $\Sigma_{11}$ is a block diagonal matrix with blocks given by $\Sigma_{\omega+\varepsilon}$ and defined by \eqref{eq:Sigmaomegavarepsilon}.
%
Moreover, $\mu_2=x(s_0,t_0)\beta$, where $x(s_0,t_0)$ is the $k$-dimensional 
covariate vector observed at time $t_0$ at site $s_{0}$, and $\Sigma_{22}=\sigma^2_\omega+\sigma^2_\varepsilon$. Finally, let us denote with $\mathbf{0}_{d\times 1}$ a $d$-dimensional row vector of zeros and with $C_\theta\left(\mathbf{\tilde h}\right)$ the correlation vector. Hence the covariance vector $\Sigma_{12}$  is defined by 
\[
\Sigma_{12}=
\begin{pmatrix}
\underbrace{\mathbf{0}_{d\times 1}}_{t=1}, \ldots,\underbrace{\sigma^{2}_{\omega}C_\theta\left(\mathbf{\tilde h}\right)}_{t=t_0},\ldots,\underbrace{\mathbf{0}_{d\times 1}}_{t=T}\\
\end{pmatrix}
'\]
where it can be seen that the $d$ non-zero elements occur at the time point $t_0$.
\subsection{Model A2}
For Model A2 it is not possible to marginalize over $U$ and thus for defining the posterior predictive distribution we refer to \eqref{eq:posteriorpredictive}  and rewrite it as follows:
%
%
\[
f\left(z(s_0,t_0)\mid Z\right)=\int f\left(z(s_0,t_0)\mid u(s_0,t_0),\Psi\right)f\left(u(s_0,t_0)\mid U, \Psi\right)f(U,\Psi\mid Z) d u(s_0,t_0) dU d\Psi
\]
where $f\left(z(s_0,t_0)\mid u(s_0,t_0), \Psi\right) \sim N_1\left(u(s_0,t_0),\sigma^2_\varepsilon\right)$. To define $f\left(u(s_0,t_0) \mid U,\Psi\right)$ we consider the following joint distribution 
\[
\left(
\begin{array}
[c]{c}%
U\\
u\left(  s_{0},t_0\right)
\end{array}
\bigg| \bigg. \Psi\right)  \sim N_{dT+1}\left[  \left(
\begin{array}
[c]{c}%
X\beta\\
x(s_0,t_0)\beta
\end{array}
\right)  ,\left(
\begin{array}
[c]{cc}%
\Sigma_U=\sigma^2_\omega C_\theta(\mathbf{h,l})& \Sigma_{12}\\
\Sigma_{12}^{\prime} & \sigma^2_\omega
\end{array}
\right)  \right]
\]
with the covariance vector $\Sigma_{12}$ given by
\[
\Sigma_{12}=\sigma^2_\omega \left(C_{\theta_1}\left(|t-t_0|\right)_{t=1,\ldots,T} \otimes C_{\theta_2}\left(\|s_i-s_0\|\right)_{i=1,\ldots,d}\right)=\sigma^2_\omega \left(C_{\theta_1}\left(\mathbf{\tilde l}\right) \otimes C_{\theta_2}\left(\mathbf{\tilde h}\right)\right).
\]
%
Now it can be easily derived that
\begin{equation}\label{eq:cond posterior A2}
u\left(s_0,t_0\right) \mid U, \Psi \sim N_1 \left(x(s_0,t_0)\beta+\frac{\Sigma_{12}^\prime C_{\theta}\left(\mathbf{ h,l}\right)^{-1}\left(U-X\beta\right)}{\sigma^2_\omega}, \sigma^2_\omega-\frac{\Sigma_{12}^\prime C_{\theta}\left(\mathbf{ h,l}\right)^{-1}\Sigma_{12}}{\sigma^2_\omega}\right).
\end{equation}
Using composition sampling, once a value from \eqref{eq:cond posterior A2} is drawn, it is possible to obtain a prediction generating a value from $f\left(z(s_0,t_0)\mid u(s_0,t_0),\Psi\right) $.

\subsection{Model A3}
For Model A3 we have that  $\mu_1=X\beta$, where $X=\{X_1,\ldots,X_T\}$ is the $(dT\times k)$ array of covariates, and $\Sigma_{11}=\Sigma_{\omega+\varepsilon}$  is the variance-covariance matrix defined in Section \ref{sec:A3}. Moreover, $\mu_2=x(s_0,t_0)\beta$ and $\Sigma_{22}=\sigma^2_\omega+\sigma^2_\varepsilon$. Finally, let us denote with $C_\theta\left(\mathbf{\tilde h},|t-t_0|\right)=C_{\theta}\left(\|s_i-s_0\|,|t-t_0|\right)_{i=1,\ldots,d}$ the $(d\times 1)$-dimensional correlation vector at time $t$ evaluated through the nonseparable correlation function $C_\theta(h,l)$ given in Eq.\eqref{covGNE}. It follows that the covariance vector $\Sigma_{12}$ is given by
\[
\Sigma_{12}=\sigma^2_\omega 
\begin{pmatrix}
C_{\theta}\left(\mathbf{\tilde h},|1-t_0|\right),\ldots,C_{\theta}\left(\mathbf{\tilde h},|t-t_0|\right),\ldots,C_{\theta}\left(\mathbf{\tilde h},|T-t_0|\right)
\end{pmatrix}^\prime
.\]

\subsection{Model B}  
For Model B Eq.\eqref{eq:posteriorpredictive} needs to be rewritten as follows in order to consider $Y$:
\begin{equation}\label{eq:posterior predictive 3}
f\left(z(s_0,t_0)\mid Z\right)=\int f\left(z(s_0,t_0)\mid y(s_0,t_0),\Psi\right)f\left(y(s_0,t_0)\mid Y, \Psi\right)f(Y,\Psi\mid Z) d y(s_0,t_0) d\Psi.
\end{equation}
As described in Section \ref{sec:B}, the process $Y$ for a given time point is constant over space so that $y(s_0,t_0)=y(t_0)$ for any $s_0$, and $y(t_0)$ is estimated using the FFBS algorithm. To draw a value from the distribution $f\left(z(s_0,t_0)\mid y(s_0,t_0), \Psi\right)$, we refer to Eq.\eqref{eq:cond posterior} with $\mu_1=X\beta+Y$ (where $X=\{X_1,\ldots,X_T\}$ and $Y=\lbrace Y_1,\ldots,Y_T\rbrace$), $\mu_2=x(s_0,t_0)\beta+y(t_0)$, $\Sigma_{22}=\sigma^2_\omega+\sigma^2_\varepsilon$ and $\Sigma_{11}$ is a block diagonal matrix with blocks given by $\Sigma_{\omega+\varepsilon}$ and defined by \eqref{eq:Sigmaomegavarepsilon}.
Finally, the covariance vector $\Sigma_{12}$ is given by
\[
\Sigma_{12}=
\begin{pmatrix}
\underbrace{\mathbf{0}_{d\times 1}}_{t=1}, \ldots,\underbrace{\sigma^{2}_{\omega}C_\theta\left(\mathbf{\tilde h}\right)}_{t=t_0},\ldots,\underbrace{\mathbf{0}_{d\times 1}}_{t=T}\\
\end{pmatrix}
'\]
as in the case of Model A1 of Section \ref{sec:A1SP}.
\subsection{Model C}
With reference to Model C, Eq.\eqref{eq:posterior predictive 3} still holds but the term $f(y(s_0,t_0) \mid Y, \Psi)$ has to be derived appropriately. To this purpose, consider the following joint distribution:
\[
\left(
\begin{array}
[c]{c}%
Y\\
y\left(  s_{0},t_0\right)
\end{array}
\bigg| \bigg. \Psi\right)  \sim N_{dT+1}\left[  \left(
\begin{array}
[c]{c}%
\rho \tilde Y\\
\rho y(s_0,t_0-1)
\end{array}
\right)  ,\left(
\begin{array}
[c]{cc}%
\Sigma_{11} & \Sigma_{12}\\
\Sigma_{12}^{\prime} & \sigma^2_\omega
\end{array}
\right)  \right]
\]
where $\tilde Y =\lbrace Y_0,\ldots,Y_{T-1}\rbrace$,  the variance-covariance matrix of $Y$ is a block diagonal matrix with blocks given by  $\Sigma_\omega=\sigma^2_\omega C_\theta\left(\mathbf{\tilde h}\right)$, 
and the covariance vector $\Sigma_{12}$ is
\[
\Sigma_{12}=
\begin{pmatrix}
\underbrace{\mathbf{0}_{d\times 1}}_{t=1}, \ldots,\underbrace{\sigma^{2}_{\omega}C_\theta\left(\mathbf{\tilde h}\right)}_{t=t_0},\ldots,\underbrace{\mathbf{0}_{d\times 1}}_{t=T}\\
\end{pmatrix}
',\]
as in the case of Model A1 and of Model B. Thus, it can be easily derived that
\begin{equation}\label{eq:cond posterior C}
y\left(s_0,t_0\right) \mid Y, \Psi \sim N_1 \left(\rho y(s_0,t_0-1)+\Sigma_{12}^\prime \Sigma_{11}^{-1}\left(Y-\rho\tilde Y\right),\sigma^2_\omega - \Sigma_{12}^\prime \Sigma_{11}^{-1} \Sigma_{12}\right).
\end{equation}

Using composition sampling for the posterior predictive distribution \eqref{eq:posterior predictive 3}, once a value from \eqref{eq:cond posterior C} is drawn, we can obtain a prediction generating a value from $f\left(z(s_0,t_0)\mid y(s_0,t_0),\Psi\right)$.


\newpage

\begin{table}
\begin{center}
\begin{tabular}{l| c c}
\hline
           &    $\psi$ & $\varphi$ \\
\hline
     Model A3-1 &      $\psi(x)=(ax^\alpha +b)/(b(ax^\alpha + 1))$ &    $\varphi(x)=\exp(-cx^\gamma)$ \\
     Model A3-2 &      $\psi(x)= (ax^\alpha + 1)^\tau$ &  $\varphi(x)=(1+cx^\gamma)^{-\nu}$ \\
\hline
\end{tabular}  
\caption{Function $\psi(x)$ and $\varphi(x)$ that define the nonseparable spatio-temporal correlation function of Model A3-1 and A3-2.}
\label{Tab:nonseparable}
\end{center}
\end{table}


\begin{footnotesize}
\begin{table}[htdp]
\begin{tabular}{@{\extracolsep{-.2cm}} l|cc|cc|cc}
\hline
 & \multicolumn{ 2}{c|}{Model A1} & \multicolumn{ 2}{c|}{Model A2} & \multicolumn{ 2}{c}{Model A3-1} \\ 
 \hline
Intercept & 3.881 & (3.815; 3.946) & 3.938	& (3.868; 4.003)  & 3.935 & (3.848; 4.021) \\ 
A & -0.155 & (-0.169; -0.140) &     -0.142	& (-0.220;	-0.062) & -0.162 & (-0.195; -0.128) \\ 
UTMX & -0.102 & (-0.124; -0.080) &  -0.110	& (-0.197;	-0.027)  & -0.112 & (-0.155; -0.069) \\ 
UTMY & -0.079 & (-0.097; -0.061) &  -0.072	&(-0.121;	-0.015)     & -0.076 & (-0.108; -0.044) \\ 
WS & -0.091 & (-0.109; -0.073) &  -0.078	&(-0.107;	-0.054) & -0.086 & (-0.101; -0.072) \\ 
HMIX & -0.041 & (-0.047; -0.035) &  -0.075	&(-0.101;	-0.046) & -0.050 & (-0.056; -0.044) \\ 
TEMP & -0.262 & (-0.309; -0.217) &   -0.103	&(-0.166;	-0.051) & -0.138 & (-0.171; -0.103) \\ 
PREC & -0.085 & (-0.107; -0.062) &  -0.105	&(-0.127;	-0.082)  & -0.090 & (-0.105;  -0.076) \\ 
EMI & 0.063 & (0.053; 0.073) &    0.083	&(0.038; 0.133)  & 0.054 & (0.031; 0.076) \\ 
\hline
 & \multicolumn{ 2}{c|}{Model A3-2} & \multicolumn{ 2}{c|}{Model B} & \multicolumn{ 2}{c}{Model C} \\ 
 \hline
Intercept &3.928 &	(3.853; 4.001) & 3.955&	(3.773; 4.119)	& 3.956	& (3.672; 	4.197)\\
A & -0.157	& (-0.187; -0.127) & -0.172 & (-0.186; -0.157) & -0.152 & (-0.187; -0.118) \\ 
UTMX & -0.108	& (-0.145; -0.071) & -0.113 & (-0.130; -0.095) & -0.092 & (-0.173; 0.001) \\ 
UTMY & -0.074	& (-0.102; -0.047) & -0.073 & (-0.086; -0.060) & -0.129 & (-0.196; -0.059) \\ 
WS & -0.090	& (-0.105; -0.076) & -0.085 & (-0.100; -0.071) & -0.032 & (-0.052; -0.012) \\ 
HMIX  & -0.050	& (-0.056; -0.044) & -0.042 & (-0.048; -0.036) & -0.024 & (-0.041; -0.007) \\ 
TEMP & -0.128	& (-0.165; -0.091) & -0.308 & (-0.349; -0.267) & -0.218 & (-0.296; -0.119) \\ 
PREC & -0.093 &	(-0.107; -0.078) & -0.090 & (-0.106; -0.074) & -0.040 & (-0.060; -0.019) \\ 
EMI & 0.058 &	(0.037; 0.079) & 0.061 & (0.050; 0.072) & 0.049 & (0.030; 0.067) \\ 
\hline
\end{tabular}
\caption{Posterior estimates (mean and 95\% credible interval in brackets) of the covariate coefficient vector $\beta$.}
\label{Tab:beta}
\end{table}
\end{footnotesize}


\begin{footnotesize}
\begin{table}[htbp]
\begin{tabular}{@{\extracolsep{-.1cm}} l|cc|cc|cc}
 \hline
 & \multicolumn{ 2}{c|}{Model A1} & \multicolumn{ 2}{c|}{Model A2} & \multicolumn{ 2}{c}{Model A3-1} \\ 
 \hline
$\sigma^2_\omega$ & 0.237	& (0.198; 0.283) & 0.247	& (0.229; 0.267) &	0.077&	(0.070;	0.083)\\
$\sigma^2_\varepsilon$ & 0.071 & (0.066; 0.077) & 0.015	&(0.013; 0.018) & 0.034 & (0.028; 0.039) \\ 
\hline
 & \multicolumn{ 2}{c|}{Model A3-2} & \multicolumn{ 2}{c|}{Model B} & \multicolumn{ 2}{c}{Model C} \\ 
 \hline
$\sigma^2_\omega$ &0.077	& (0.071; 0.084)	&0.063	&(0.053; 0.073)	&0.950	&(0.822;	1.091)\\
$\sigma^2_\varepsilon$ & 0.034	& (0.029; 0.040) & 0.051 & (0.042; 0.060) & 0.013 & (0.011; 0.015) \\ 
$\sigma^2_\eta$ & -- &  -- & 0.065 & (0.052; 0.081) & --&  -- \\ 
\hline
\end{tabular}
\caption{Posterior estimates (mean and 95\% credible interval in brackets) of the variance parameters.}
\label{Tab:variances}
\end{table}
\end{footnotesize}


\begin{footnotesize}
\begin{table}[htbp]
\begin{tabular}{@{\extracolsep{-.1cm}} l|cc|cc}
 \hline
 & \multicolumn{ 2}{c|}{Model A1} & \multicolumn{ 2}{c}{Model A2} \\ 
 \hline
$\theta$ & 0.0033	&(0.0025;	0.0042)	& --	& --	\\
$\theta_1$ & -- & -- & 0.492 &	(0.473; 0.511)\\ 
$\theta_2$ & -- & --  & 0.032	& (0.029; 0.035)  \\ 
\hline
& \multicolumn{ 2}{c|}{Model B} & \multicolumn{ 2}{c}{Model C}\\
\hline
$\theta$&0.050 & (0.040; 0.061)&	0.0022&	(0.0019; 0.0024)\\
$\rho$ & 0.831	&(0.738;	 0.919)	& 0.654	&(0.629;	0.677)\\
\hline
\end{tabular}
\caption{Posterior estimates (mean and 95\% credible interval in brackets) of the correlation function parameters for Model A1, A2, B and C.}
\label{Tab:correlation}
\end{table}
\end{footnotesize}


\begin{footnotesize}
\begin{table}[htbp]
\begin{tabular}{@{\extracolsep{-.1cm}} l|cc|cc}
 \hline
 & \multicolumn{ 2}{c|}{Model A3-1} & \multicolumn{ 2}{c}{Model A3-2} \\ 
 \hline
$a$&	0.058&	(0.045; 0.074)&	4.088	& (3.533;  4.708)\\
$\alpha$&	0.736&	(0.650; 0.831)	& 0.800 & (0.743; 0.859) 	\\
$b$	& 0.047 & (0.036;	 0.058)	& -- & --\\
$c$	&0.549	& (0.448; 0.659)	& 0.982 & (0.823; 1.158) \\
$\gamma$&	 0.110&	(0.089; 0.132)	&0.184 & (0.156; 0.212) \\
$\nu$	&-- &	--		&0.801 & (0.743; 0.864) 	\\
$\tau$	&-- &	--	&0.4756 & (0.434; 0.517) 	\\
\hline
\end{tabular}
\caption{Posterior estimates (mean and 95\% credible interval in brackets) of the correlation parameters for Model A3-1 and A3-2.}
\label{Tab:nonseparable_par}
\end{table}
\end{footnotesize}


\begin{footnotesize}
\begin{sidewaystable}[htdp]
\begin{tabular}{@{\extracolsep{-0cm}} l|cccccc}
\hline
& \multicolumn{6}{c}{Model}\\
Criteria & A1 & A2 & A3-1 & A3-2 & B & C\\
\hline
N. of parameters ($\beta$'s excluded) & 3 &  4 & 7 & 8 & 5 & 4\\
-- estimated by MH & 3 & 2& 7 & 8 & 4 & 2\\ 
Size of the biggest matrix to be inverted &  $(d \times d)$ &  $(T \times T)$ &  $(dT \times dT)$ &  $(dT \times dT)$ &  $(d \times d)$ &  $(d \times d)$\\
\hdashline
Estimation time (sec per iteration) &0.014 &  3.992 & 45.299  & 47.466 & 0.063 &0.360 \\
Prediction time (sec per iteration) &0.026 & 27.288  &  17.733 & 20.367 & 0.024 & 0.038\\
\hdashline
Prediction capability &** & *  &** & ** & ** & ***\\
\hline 
\end{tabular}
\caption{Performance and complexity of the models.}
\label{Tab:comparison}
\end{sidewaystable}
\end{footnotesize}


\end{document}